\documentclass[epj,nopacs]{svjour}
\usepackage[T1]{fontenc}
\usepackage{amsmath,amssymb,mathtools,bm}
\usepackage{graphicx,booktabs,array,placeins}
\usepackage[numbers,sort&compress]{natbib}
\usepackage{microtype}
\usepackage[colorlinks=true,linkcolor=black,citecolor=black,urlcolor=black]{hyperref}
\newcommand{\dd}{\mathrm d}
\newcommand{\mpl}{M_{\mathrm{Pl}}}
\newcommand{\bta}{\Delta\beta}
\newcommand{\Msun}{M_\odot}
\providecommand{\prd}{Phys. Rev. D}
\providecommand{\doi}[1]{\href{https://doi.org/#1}{doi:#1}}
\begin{document}
\title{Axial gravitational-wave propagation and mass--gravity identifiability in conformal scalar--tensor lensing}
\titlerunning{Axial-wave propagation in conformal scalar--tensor lensing}
\author{Z. F. Gao\inst{1} \and W. Q. Ma\inst{1,2} \and H. Wang\inst{3} \and Z. J. Cao\inst{4,5} \and N. Wang\inst{1} \and G. Q. Jin\inst{6} \and Luiz C. Garcia de Andrade\inst{7}}
\authorrunning{Z. F. Gao et al.}
\institute{State Key Laboratory of Radio Astronomy and Technology, Xinjiang Astronomical Observatory, Chinese Academy of Sciences, Urumqi 830011, China
\and University of Chinese Academy of Sciences, Beijing 100049, China
\and Shanxi Province Intelligent Optoelectronic Sensing Application Technology Innovation Center, Yuncheng University, Yuncheng, China
\and Institute for Frontiers in Astronomy and Astrophysics, Beijing Normal University, Beijing 102206, China
\and School of Physics and Astronomy, Beijing Normal University, Beijing 100875, China
\and College of Mechanical and Electrical Engineering, Tarim University, Alar 843300, China
\and Instituto de F\'isica, Universidade do Estado do Rio de Janeiro, Rio de Janeiro 20550-013, Brazil}
\mail{zhifugao@xao.ac.cn; wanghuiycu@163.com}
\date{}
\abstract{We investigate tensor-wave propagation in a specified massless, unscreened conformal scalar--tensor theory using a static Einstein--scalar exterior matched to a weakly self-gravitating matter lens. The odd-parity gravitational master potential has a scalar-charge correction at radial order $R^{-4}$ for fixed Einstein-frame mass. Its next $R^{-5}$ coefficient depends on $\ell(\ell+1)-9$, reversing sign between the quadrupole and higher multipoles; we test the expansion against the exact exterior potential. We connect this result to the locally normalized post-Newtonian light-deflection parameter and distinguish the Schwarzschild baseline at fixed Einstein-frame mass from that at fixed measured matter mass. A finite-radius WKB expansion reveals cancellation of the leading frequency-dependent potential moment for the quadrupole at fixed multipole, whereas a separate fixed-impact-parameter limit recovers an achromatic geometrical-optics action and a subleading axial contribution. Independent null-geodesic bending and numerical quadrature check these limits. For a point lens, achromatic observables constrain an effective mass rather than the lens mass and scalar coupling separately. An illustrative two-image Fisher calculation quantifies differential-phase identifiability after arrival-time and constant-phase nuisance parameters are released. The phenomenological phase experiment is separated from action-derived propagation quantities; model-specific dispersive observations require a complete matched image transfer function.}
\maketitle

\section{Introduction}\label{sec:intro}
The observation of gravitational waves (GWs) from compact binaries has opened a direct route to testing relativistic gravity in the radiative and strong-field regimes \citep{Abbott2016,Abbott2019PRL,Abbott2021}. The multi-messenger event GW170817 has also established stringent low-redshift constraints on the speed of tensor waves \citep{Abbott2017,Ezquiaga2017}. These achievements do not, however, determine the response of all additional degrees of freedom that arise in gravity theories beyond general relativity (GR). Scalar--tensor theories, including Horndeski gravity and its luminal subclasses, offer a well-defined setting in which scalar backgrounds, source generation and wave propagation must be treated consistently \citep{BransDicke1961,Horndeski1974,Clifton2012,Kobayashi2011,Bettoni2013,Bellini2014,Saltas2014}. In particular, cosmological propagation functions describing tensor-speed and damping modifications cannot by themselves determine the local wave operator in the strongly inhomogeneous field of an intervening lens \citep{Lombriser2016,Belgacem2018,Nishizawa2018,Ezquiaga2018Frontiers}.

Gravitational lensing adds a different observational configuration: multiple images are associated with the same source but follow different paths through the foreground potential. In geometric optics, their relative arrival time, magnification and Morse phase depend on the lens geometry, whereas diffraction and finite-frequency corrections require wave-optics information \citep{Takahashi2003,Dai2017,Bartelmann2010,Ezquiaga2020PRD,Ezquiaga2021PRD,Goyal2023PRD,Savastano2019PRD}. This distinction matters for future ET and LISA observations, whose sensitivities and source populations make resolved or overlapping lensed signals scientifically relevant \citep{Hild2011,AmaroSeoane2017,Robson2019,Auclair2023,Grespan2023,Finke2021}. It is nevertheless incorrect to infer detectability from the lensing time delay alone: both images must fall within the observing window and satisfy signal-to-noise and selection requirements \citep{LiMao2018}. The use of a simple point lens is best understood as a transparent benchmark rather than a realistic cluster or galaxy mass model.

Theoretical progress has made the relevant physical distinction sharper. The short-wave expansion of scalar and tensor perturbations can be performed on general backgrounds, including nontrivial mixing and corrections beyond leading geometrical optics \citep{Isaacson1968,Dalang2021,Garoffolo2020}. Menadeo and Zumalac\'arregui derived lens-induced beyond-geometric-optics effects for a Brans--Dicke benchmark and emphasized their severe suppression under existing weak-field bounds \citep{Menadeo2025}. More recently, Menadeo, Giardino and Zumalac\'arregui studied dispersion on inhomogeneous backgrounds, including screened and non-minimally coupled settings \citep{MenadeoGiardino2026}; related approaches examine wave-optics lensing in modified gravity \citep{GaroffoloTasinato2026}. A recent EPJC perturbation analysis with angularly structured scalar hair demonstrates that axial--scalar closure depends on the parity content of the matter background, rather than on spherical symmetry of the stress tensor alone \citep{Nashed2026Odd}. Our radial-scalar exterior belongs to the distinct parity-decoupled case. Consequently, a generic $1/f$ phase, a principal null cone shared by modes, or a phenomenological image-difference parameter is not by itself a new Horndeski prediction. A model-specific result must follow from the chosen action, a physically admissible lens background and a consistently defined source-to-detector transfer calculation.

A central difficulty is that several apparently similar calculations answer distinct questions. The cosmological effective-field-theory description is organized by homogeneous-background functions \citep{Bellini2014,Lagos2018}, while local lensing depends on gradients and boundary conditions. A scalar flux equation may display a $1/f$ WKB correction without representing a physical tensor polarization. Likewise, a fixed-$\ell$ radial partial wave is not automatically a finite-impact-parameter lensed wavepacket, and a potential difference is not automatically an observable GR-subtracted phase. The spherical Einstein--massless-scalar exterior provides a controlled laboratory in which odd-parity gravitational perturbations can be separated from the even-parity scalar sector \citep{Karlovini2002,ChowdhuryBanerjee2020}. Its limitations must also be stated: an isolated scalar-charged exterior is not a self-consistent description of every extended astrophysical lens, and a constant-scalar vacuum black-hole branch need not generate the postulated anomalous source phase \citep{Will2014}.

Here we focus on a specified, massless and unscreened conformal scalar--tensor subclass, not an unconstrained four-function Horndeski model. The objective is to determine which pieces of a proposed GW-lensing test are actually implied by this action and which remain conditional diagnostics. Our contributions are fourfold. First, we connect the Jordan-frame Cavendish coupling, the Einstein-frame mass normalization and the spherical scalar exterior, construct the physical odd-parity tensor master potential, and derive its multipole-dependent $R^{-5}$ correction, verified against the exact potential. Second, we distinguish its fixed-multipole radial WKB moment from the fixed-impact-parameter eikonal limit, checking the latter against an independent null-geodesic bending calculation. Third, we examine the model's leading achromatic lens geometry, the degeneracy of lens mass with a post-Newtonian parameter, and the conditions needed to observe both images. Fourth, using an explicitly phenomenological common-plus-differential $1/f$ phase, we compare reduced five-parameter and image-time/phase-marginalized nine-parameter Fisher calculations. The last calculation is a measurement-identifiability diagnostic, not a forecast for the fundamental scalar coupling \citep{Cutler1994,Vallisneri2008,Carson2020}.

The calculation establishes an exact odd-parity exterior sector, two complementary high-frequency limits and mass--gravity degeneracy relations. The separate phase-coefficient experiment supplies a test of statistical identifiability for future analyses that include both image transfer functions and complete waveform information.

The paper is organized in seven sections. Section~\ref{sec:setting} develops the covariant perturbation framework; section~\ref{sec:reduced} derives the axial exterior result and compares the two WKB limits with the scalar surrogate. Section~\ref{sec:pnlens} analyzes the weak-field lens map and mass normalization, section~\ref{sec:two} examines conditional two-image phase identifiability, and section~\ref{sec:timedelay} addresses the observing-window condition. Section~\ref{sec:discussion} summarizes the conclusions, followed by two technical appendices.

\noindent\textbf{Notation and provenance.} The Jordan scalar is $\phi$ and the canonical Einstein-frame scalar is $\chi$. The dimensionless coupling $Q$ is a parameter of the specified action, not a measured phenomenological phase. $G_{\rm bare}$ denotes the bare tensor-sector constant; $G_N$ is the locally measured Cavendish constant. For a Jordan-frame matter lens of mass $M_L$, $r_g=G_NM_L/c^2$, while $m_E$ is the independent Einstein-frame ADM mass expressed as a length. The FJNW exterior uses areal radius $R$, coordinate $r$, and parameters $(B,\zeta)$. $r_*$ is its optical radial coordinate, $\ell\ge 2$ the multipole index, $\omega$ the angular frequency in Einstein-frame geometrized units and $b=\sqrt{\ell(\ell+1)}/\omega$ the high-multipole impact parameter. $f$ denotes the observed frequency in hertz. $D_{\rm OL},D_{\rm OS},D_{\rm LS}$ are angular-diameter distances and $D_{\rm lum}$ is the waveform luminosity distance. We specify the mass- and endpoint-matching conventions whenever making GR comparisons \citep{BransDicke1961,Will2014,Karlovini2002}.

\section{Covariant framework and tensor propagation}\label{sec:setting}
We work with the Jordan-frame Horndeski action and minimally coupled matter, using $X=-g^{\mu\nu}\nabla_\mu\phi\nabla_\nu\phi/2$ and metric signature $(-,+,+,+)$:
\begin{equation}
S=\int\dd^4x\sqrt{-g}\sum_{i=2}^5{\cal L}_i+S_m[g_{\mu\nu},\psi_m].
\end{equation}
The usual four densities, involving $G_i(\phi,X)$, define the covariant model \citep{Horndeski1974,Kobayashi2011}. The familiar Bellini--Sawicki effective functions $(\alpha_K,\alpha_B,\alpha_M,\alpha_T)$ are defined on a homogeneous FLRW solution, not on a generic lens background \citep{Bellini2014}. In particular,
\begin{align}
M_*^2 &=2\bigl(G_4-2XG_{4X}+XG_{5\phi}-H\dot{\bar\phi}XG_{5X}\bigr),\\
\alpha_M&=H^{-1}\dd(\ln M_*^2)/\dd t,\\
M_*^2\alpha_T&=2X\left[2G_{4X}-2G_{5\phi}-(\ddot{\bar\phi}-H\dot{\bar\phi})G_{5X}\right].
\end{align}
The other two background functions constrain kineticity and braiding, but none of these quantities alone specifies the local lens profile or wave operator. We therefore do not introduce a general ``$\Upsilon_L(\alpha_i)$'' lens-response formula.

Locally, write $\phi=\bar\phi+\delta\phi_L+\varphi$ and $\dd s^2=-(1+2\Phi)c^2\dd t^2+(1-2\Psi)\dd\bm x^2$. For a fully coupled system after constraints have been eliminated, the propagating variables schematically obey
\begin{equation}
\begin{aligned}
[ K_{IJ}^{\mu\nu}\nabla_\mu\nabla_\nu
+A_{IJ}^{\mu}\nabla_\mu+U_{IJ}]\psi^J&=0,\\
\psi^I&=(h_+,h_\times,\varphi).
\end{aligned}
\label{eq:fullop}
\end{equation}
The WKB principal symbol $P_{IJ}=K_{IJ}^{\mu\nu}k_\mu k_\nu$ governs the characteristic and polarization branches, while transport and background gradients enter at the next orders \citep{Isaacson1968,Garoffolo2020,Ezquiaga2020PRD}. The local matrices in \eqref{eq:fullop} must be computed from the same action and background to obtain a general Horndeski lens-induced dispersion. We do not claim to have completed that computation.

\subsection{Coupled metric--scalar perturbations}\label{sec:covariant}
The restricted conformal model in equation~\eqref{eq:model} admits an exact Einstein-frame rewriting, which permits a check of terms absent from a longitudinal propagation equation. Defining $g^E_{\mu\nu}=e^{2Q\phi/\mpl}g^J_{\mu\nu}$ and $\chi(\phi)$ by
\begin{equation}
 \left(\frac{\dd\chi}{\dd\phi}\right)^2=e^{-2Q\phi/\mpl}+6Q^2,
 \qquad A(\chi)=e^{-Q\phi(\chi)/\mpl},
 \label{eq:efredef}
\end{equation}
the action outside the matter source is
\begin{equation}
 S_{E,\mathrm{ext}}=\int\dd^4x\sqrt{-g_E}\left[\frac{\mpl^2}{2}\mathcal R_E
 -\frac12g_E^{\mu\nu}\partial_\mu\chi\partial_\nu\chi\right].
 \label{eq:efaction}
\end{equation}
Matter, where present, couples to $A^2(\chi)g^E_{\mu\nu}$. The exact exterior field equations are
\begin{equation}
 R^E_{\mu\nu}=\mpl^{-2}\nabla_\mu\chi\nabla_\nu\chi,
 \qquad \Box_E\chi=0.
 \label{eq:backgroundE}
\end{equation}
Let $(g^E_{\mu\nu},\chi_0)$ solve equation~\eqref{eq:backgroundE} and consider
$g^E_{\mu\nu}\mapsto g^E_{\mu\nu}+h^E_{\mu\nu}$ and $\chi_0\mapsto\chi_0+\eta$.
Direct variation gives the following unfixed-gauge, covariant, coupled equations in the exterior:
\begin{align}
\delta R^E_{\mu\nu}[h^E]={}&\mpl^{-2}
(\nabla_\mu\chi_0)(\nabla_\nu\eta)\notag\\
&+\mpl^{-2}(\nabla_\nu\chi_0)(\nabla_\mu\eta),
\label{eq:linmetric}\\
\Box_E\eta={}&h_E^{\mu\nu}\nabla_\mu\nabla_\nu\chi_0\notag\\
&+\left(\nabla_\mu h_E^{\mu\nu}
-\tfrac12\nabla^\nu h_E\right)\nabla_\nu\chi_0.
\label{eq:linscalar}
\end{align}
where $h_E=g_E^{\mu\nu}h^E_{\mu\nu}$ and all derivatives refer to the background Einstein metric. The Ricci variation is explicitly
\begin{equation}
\begin{aligned}
2\delta R^E_{\mu\nu}[h^E]={}&\nabla_\rho\nabla_\mu h_E{}^\rho{}_\nu
+\nabla_\rho\nabla_\nu h_E{}^\rho{}_\mu\\
&-\Box_Eh^E_{\mu\nu}-\nabla_\mu\nabla_\nu h_E.
\end{aligned}
\label{eq:riccivar}
\end{equation}
Equations~\eqref{eq:linmetric}--\eqref{eq:linscalar} retain scalar--metric mixing, curved-background derivatives and their associated curvature commutators; no transverse derivatives have been discarded. They provide action-derived local equations but not yet a gauge-fixed physical tensor/scalar transfer matrix or a lensing phase. Under an infinitesimal background diffeomorphism, $h^E_{\mu\nu}\mapsto h^E_{\mu\nu}+2\nabla_{(\mu}\xi_{\nu)}$ and $\eta\mapsto\eta+\xi^\rho\nabla_\rho\chi_0$; this freedom must be fixed or handled by a gauge-invariant construction before interpreting the metric components as observable polarizations. In particular, a coordinate-gauge choice, physical polarization extraction, incident-state specification, boundary conditions and consistent GR subtraction are required before a phase can be assigned to an observable tensor mode.

\paragraph{Harmonic-gauge propagation and characteristics.}\label{sec:harmonic}
One can go beyond the unreduced equations without assuming a longitudinal flux-weighted tensor channel. Impose the local background-covariant de Donder condition on the Einstein-frame metric perturbation,
\begin{equation}
\nabla^\mu\bar h^E_{\mu\nu}=0,\qquad
\bar h^E_{\mu\nu}=h^E_{\mu\nu}-\frac12g^E_{\mu\nu}h_E.
\label{eq:harmonic}
\end{equation}
Commuting covariant derivatives in equation~\eqref{eq:riccivar} and using equation~\eqref{eq:harmonic} gives the coupled, gauge-fixed but not constraint-reduced exterior system
\begin{align}
-\tfrac12\Box_E h^E_{\mu\nu}
-R^E_{\mu\rho\nu\sigma}h_E^{\rho\sigma}\notag\\
+R^E_{\rho(\mu}h_E{}^\rho{}_{\nu)}
&=\frac{2}{\mpl^2}
(\nabla_{(\mu}\chi_0)(\nabla_{\nu)}\eta),
\label{eq:gfh}\\
\Box_E\eta-h_E^{\mu\nu}\nabla_\mu\nabla_\nu\chi_0&=0.
\label{eq:gfscalar}
\end{align}
The right-hand side of equation~\eqref{eq:gfh} has only one derivative of the perturbation $\eta$; the background curvature terms have none. The scalar equation contains only an algebraic perturbation term in this gauge. Accordingly, for the rapidly varying amplitudes $(a_{\mu\nu},b)e^{i\theta/\epsilon}$, the two-derivative principal equations are
\begin{equation}
(g_E^{\rho\sigma}k_\rho k_\sigma)a_{\mu\nu}=0,
\qquad (g_E^{\rho\sigma}k_\rho k_\sigma)b=0,
\quad k_\mu=\nabla_\mu\theta,
\label{eq:characteristic}
\end{equation}
with the leading harmonic constraints $k^\mu(a_{\mu\nu}-g^E_{\mu\nu}a/2)=0$. After accounting for residual diffeomorphisms and constraints, the two tensor modes and one scalar therefore share the Einstein-metric null cone at principal order. Conformal rescaling preserves this null cone in the Jordan frame. This is a model-specific absence of geometric-optics speed splitting, not an absence of finite-frequency dispersion or mode conversion. The Riemann term and the scalar-gradient coupling in equation~\eqref{eq:gfh} remain essential when the WKB hierarchy is continued to subleading orders, despite the transverse projection of their leading derivative part in section~\ref{sec:tttransport}. The source's interior, the scattering boundary conditions and the tensor-polarization projection are not fixed by this local characteristic statement.

\paragraph{Screen-transverse transport.}\label{sec:tttransport}
The derivative coupling on the right-hand side of equation~\eqref{eq:gfh} can be tested more directly than by inspecting the characteristic determinant. Along a null characteristic $k_\mu=\nabla_\mu\theta$, choose a locally parallel-transported screen basis of symmetric transverse--traceless tensors $e_A^{\mu\nu}$, $A=+,\times$, satisfying
\begin{equation}
 k_\mu e_A^{\mu\nu}=0,\qquad g^E_{\mu\nu}e_A^{\mu\nu}=0,
 \qquad k^\rho\nabla_\rho e_A^{\mu\nu}=0.
 \label{eq:screenbasis}
\end{equation}
For the leading shortwave fields, take amplitudes varying on the background scale and write
\begin{equation*}
\begin{aligned}
h^E_{\mu\nu}&=\sum_A\mathcal A_A e^A_{\mu\nu}
 e^{i\theta/\epsilon}+\text{non-TT components},\\
\eta&=\mathcal B e^{i\theta/\epsilon}.
\end{aligned}
\end{equation*}
The potentially leading scalar source projected on $e_A$ obeys the algebraic identity
\begin{equation}
 e_A^{\mu\nu}\nabla_{(\mu}\chi_0\nabla_{\nu)}\eta
 =\frac{i}{\epsilon}{\cal B}e^{i\theta/\epsilon}
 e_A^{\mu\nu}(\nabla_\mu\chi_0)k_\nu+O(1)
 =O(1).
 \label{eq:ttcancel}
\end{equation}
Thus the derivative mixing term has no direct screen-TT projection at order $\epsilon^{-1}$. Projecting the same order of the wave operator on the left of equation~\eqref{eq:gfh}, and taking the screen basis to be parallel transported along the ray, yields the standard geometric-optics law
\begin{equation}
 2k^\rho\nabla_\rho{\cal A}_A+(\nabla_\rho k^\rho){\cal A}_A=0,
 \qquad \nabla_\rho(|{\cal A}_A|^2k^\rho)=0.
 \label{eq:ttflux}
\end{equation}
This step assumes that the leading incident metric perturbation has been projected onto the physical screen-TT subspace and that the same eikonal branch is followed smoothly; it is not a global gauge-invariant construction of the detected polarization. To locate the first omitted terms, write $h^E_{\mu\nu}=e^{i\theta/\epsilon}(a^{(0)}_{\mu\nu}+\epsilon a^{(1)}_{\mu\nu}+\cdots)$ and $\eta=e^{i\theta/\epsilon}(b_0+\epsilon b_1+\cdots)$. Define $\mathcal D_k z\equiv2k^\rho\nabla_\rho z+(\nabla_\rho k^\rho)z$ and the curvature endomorphism
\begin{equation}
(\mathcal U a)_{\mu\nu}\equiv -R^E_{\mu\rho\nu\sigma}a^{\rho\sigma}+R^E_{\rho(\mu}a^\rho{}_{\nu)}.
\label{eq:curvop}
\end{equation}
The scalar equation at order $\epsilon^{-1}$ implies $\mathcal D_k b_0=0$; if no leading scalar wave enters, $b_0$ therefore remains zero along a smooth ray. The scalar equation at order $\epsilon^0$ then has the first inhomogeneous source,
\begin{equation}
 i\mathcal D_k b_1=a^{(0)\mu\nu}\nabla_\mu\nabla_\nu\chi_0 \quad (b_0=0).
 \label{eq:scalarborn}
\end{equation}
For this tensor-only incident condition, the screen projection of the metric equation at the same order reads
\begin{equation}
 e_A^{\mu\nu}\!\left[-\frac{i}{2}\mathcal D_k a^{(1)}_{\mu\nu}
 -\frac12\Box_E a^{(0)}_{\mu\nu}+(\mathcal U a^{(0)})_{\mu\nu}\right]=0.
\label{eq:ttbgo}
\end{equation}
The apparently leading $\nabla\chi_0\,k\,b_1$ source vanishes after transverse projection just as in equation~\eqref{eq:ttcancel}. This does not make equation~\eqref{eq:ttbgo} a completed physical phase calculation: the transport of all physical/gauge amplitudes, the curvature and polarization projections, ray boundary conditions and the matching GR baseline remain to be evaluated. A nonzero leading incident scalar $b_0$ instead adds a projected $\nabla\chi_0\,\nabla b_0$ source. These WKB order identities are consistent with geometric-optics studies of luminal scalar--tensor propagation \citep{Dalang2021,Garoffolo2020} and are not claimed as a new universal absence of mode interaction.

The order of the retained curvature terms can be checked directly. Let $r_g\equiv G_NM_L/c^2$ and use the same weak-field scalar profile as equation~\eqref{eq:charge}. Since $\dd\chi/\dd\phi|_{\phi=0}=\sqrt{1+6Q^2}$,
\begin{equation}
 \frac{\chi_0-\chi_\infty}{\mpl}
 =s_Q\frac{r_g}{r}+O(r_g^2/r^2),\qquad
 s_Q\equiv\frac{2Q\sqrt{1+6Q^2}}{1+8Q^2}.
 \label{eq:chiweak}
\end{equation}
Using the trace of equation~\eqref{eq:backgroundE}, the leading exterior Einstein-frame Ricci scalar is
\begin{equation}
 \mathcal R_E=\frac{(\nabla\chi_0)^2}{\mpl^2}
 =\frac{4Q^2(1+6Q^2)}{(1+8Q^2)^2}\frac{r_g^2}{r^4}
 +O(r_g^3/r^5).
 \label{eq:ricciscale}
\end{equation}
For comparison, the flat-space three-dimensional scalar surrogate later defined in equation~\eqref{eq:v3d} contains the non-distributional term $V_{3\mathrm D}=\varepsilon_Q^2r_g^2/(4r^4)$ on the same exterior ray. The ratio of these two different local quantities is
\begin{equation}
\begin{aligned}
 \frac{\mathcal R_E}{V_{3\mathrm D}}&=\frac{1+6Q^2}{Q^2}+O(r_g/r)\\
 &=Q^{-2}+6+O(r_g/r),\quad Q\ne0.
\end{aligned}
 \label{eq:curvatureorder}
\end{equation}
Thus the scalar-sourced background Ricci curvature starts at $O(Q^2)$ whereas the squared-gradient surrogate potential starts at $O(Q^4)$ for small $Q$. This is not a calculation of an $O(Q^2)$ observable phase: curvature contractions, polarization projections and scalar--metric source terms may cancel or combine after solving the complete coupled system. It is, however, an action-derived demonstration that the surrogate alone cannot establish the leading coupling order of the physical gravitational-wave effect.

\section{Axial perturbations and propagation diagnostics}\label{sec:reduced}
The calculation below differs from the curvature-coupled phenomenological wave-optics model of \citep{GaroffoloTasinato2026}: its axial potential follows from the specified conformal action and an exact radial-scalar exterior. It also addresses a different limit from the coupled beyond-geometrical-optics propagation treatments of \citep{Menadeo2025,MenadeoGiardino2026}. Our independent analytic result is the \emph{mass- and radius-matched} $R^{-5}$ coefficient in the odd-parity potential, together with its multipole sign change; the radial WKB, geodesic and identifiability calculations test the interpretation of that result. The local potential, the fixed-$\ell$ WKB action, and a fully matched two-image transfer function are distinct quantities. Thus the new asymptotic coefficient is presented as a reproducible exterior tensor-mode prediction, not as a detector-level phase measurement.

\subsection{Model and scalar-charge normalization}
We choose the luminal conformal subclass
\begin{equation}
G_2=X,\quad G_3=G_5=0,\quad G_4=\frac{\mpl^2}{2}\exp\!\left(\frac{2Q\phi}{\mpl}\right),
\label{eq:model}
\end{equation}
with matter minimally coupled to the Jordan metric. It has $M_*^2=\mpl^2e^{2Q\bar\phi/\mpl}$, $\alpha_T=0$, $\alpha_M=2Q\dot{\bar\phi}/(H\mpl)$, and $\alpha_B=-\alpha_M$. We take the local asymptotic field to be $\bar\phi_L=0$ by choice of normalization, work to first order in the weak-field potential $r_g/r\ll1$, and assume an unscreened nonrelativistic matter lens (not an isolated black-hole lens).

Following standard scalar--tensor weak-field normalization \citep{Clifton2012}, the Einstein-frame transformation in equation~\eqref{eq:efredef} gives $\dd\chi/\dd\phi|_{\bar\phi_L=0}=\sqrt{1+6Q^2}$ and $A(\chi)=e^{-Q\phi(\chi)/\mpl}$. The dimensionless weak-field scalar coupling is
\begin{equation*}
\alpha_0\equiv\mpl\frac{\dd\ln A}{\dd\chi}
 =-\frac{Q}{\sqrt{1+6Q^2}}.
\end{equation*}
In the massless, unscreened weak-field limit, the tensor Newton constant $G_{\rm bare}$ and locally measured Cavendish constant obey
\begin{equation}
G_N=G_{\rm bare}(1+2\alpha_0^2)
 =G_{\rm bare}\frac{1+8Q^2}{1+6Q^2}.
\label{eq:newton}
\end{equation}
The linearized scalar profile is
\begin{equation*}
 \frac{\delta\phi_L}{\mpl}=
 \frac{2Q}{1+6Q^2}\frac{G_{\rm bare}M_L}{c^2r}.
\end{equation*}
In terms of $r_g\equiv G_NM_L/c^2$ this becomes
\begin{equation}
\begin{aligned}
\frac{\delta\phi_L}{\mpl}
&=\frac{2Q}{1+8Q^2}\frac{r_g}{r}+O(r_g^2/r^2),\\
\delta\ln M_*^2&=\varepsilon_Q\frac{r_g}{r},\quad
\varepsilon_Q=\frac{4Q^2}{1+8Q^2}.
\end{aligned}
\label{eq:charge}
\end{equation}
The replacement of $G_{\rm bare}$ by $G_N$ must not be omitted when taking a finite-$Q$ upper envelope. This normalisation check does not by itself establish a complete coupled propagation phase.

\subsection{Odd-parity gravitational master equation}\label{sec:axial}
A controlled, genuinely tensorial reduction is available if the exterior of the matter lens is static and spherically symmetric. For clarity, use Einstein-frame geometric units $G_E=c=1$ and distinguish the Einstein-frame ADM mass in length units, $m_E$, from the locally measured Jordan-frame lens scale $r_g=G_NM_L/c^2$. The exterior scalar equations \eqref{eq:backgroundE} admit the Fisher--Janis--Newman--Winicour (FJNW) family \citep{ChowdhuryBanerjee2020}:
\begin{align}
\dd s_E^2={}&-f^\zeta\dd t^2+f^{-\zeta}\dd r^2\notag\\
&+r^2f^{1-\zeta}\dd\Omega^2,\\
f(r)={}&1-\frac{B}{r},\quad B=\frac{2m_E}{\zeta},\notag\\
R(r)={}&r f^{(1-\zeta)/2},\quad0<\zeta\le1,
\label{eq:jnwmetric}\\
\frac{\chi_0-\chi_\infty}{\mpl}={}&\pm
\sqrt{\frac{1-\zeta^2}{2}}\ln f(r).
\label{eq:jnwscalar}
\end{align}
Here $R$ is the circumferential (areal) radius, and $r>B$ is the exterior coordinate range. The formal continuation to $r=B$ is singular for $\zeta<1$; it is not treated as the interior of a physical galactic or stellar lens. The use of this background for an extended matter lens requires that the actual ray pass through its matter-free exterior and that an appropriate interior solution be matched at the source surface.

Because the background scalar is even under parity, a first-order odd-parity (axial) gravitational perturbation has no odd scalar-field partner. For $\ell\ge2$ the gauge-invariant axial master function obeys the known static-spherical reduction \citep{Karlovini2002,ChowdhuryBanerjee2020},
\begin{equation}
\frac{\dd^2 Z_{\ell}^{\rm odd}}{\dd r_*^2}+
\bigl[\omega^2-V_{\ell}^{\rm odd}(R)\bigr]Z_{\ell}^{\rm odd}=0,
\qquad \frac{\dd r_*}{\dd r}=f^{-\zeta},
\label{eq:oddmaster}
\end{equation}
with (using the Misner--Sharp mass $m(R)=\tfrac R2[1-g_E^{RR}]$)
\begin{equation}
V_{\ell}^{\rm odd}=f^\zeta\left[\frac{\ell(\ell+1)}{R^2}-\frac{6m(R)}{R^3}
 +4\pi\bigl(\rho_E-p_{r,E}\bigr)\right].
\label{eq:oddgeneral}
\end{equation}
For a canonical, static, radial massless scalar, the radial pressure equals the energy density, $p_{r,E}=\rho_E$. Inserting $R(r)$ then gives the explicit physical axial master potential
\begin{equation}
V_{\ell}^{\rm odd}(r)=\frac{f^{2\zeta-1}}{r^2}
\left[\ell(\ell+1)-\frac{3\zeta B}{r}
+\frac{3(1-\zeta)^2B^2}{4r^2f}\right].
\label{eq:oddexplicit}
\end{equation}
For $\zeta=1$ and $B=2m_E$, equation~\eqref{eq:oddexplicit} reduces exactly to the Schwarzschild Regge--Wheeler potential $(1-2m_E/R)[\ell(\ell+1)/R^2-6m_E/R^3]$. This physical odd-parity tensor channel retains radial and angular curvature, unlike the longitudinal scalar surrogate in equation~\eqref{eq:oneD}.

\begin{figure}[!t]\centering
\includegraphics[width=.89\columnwidth]{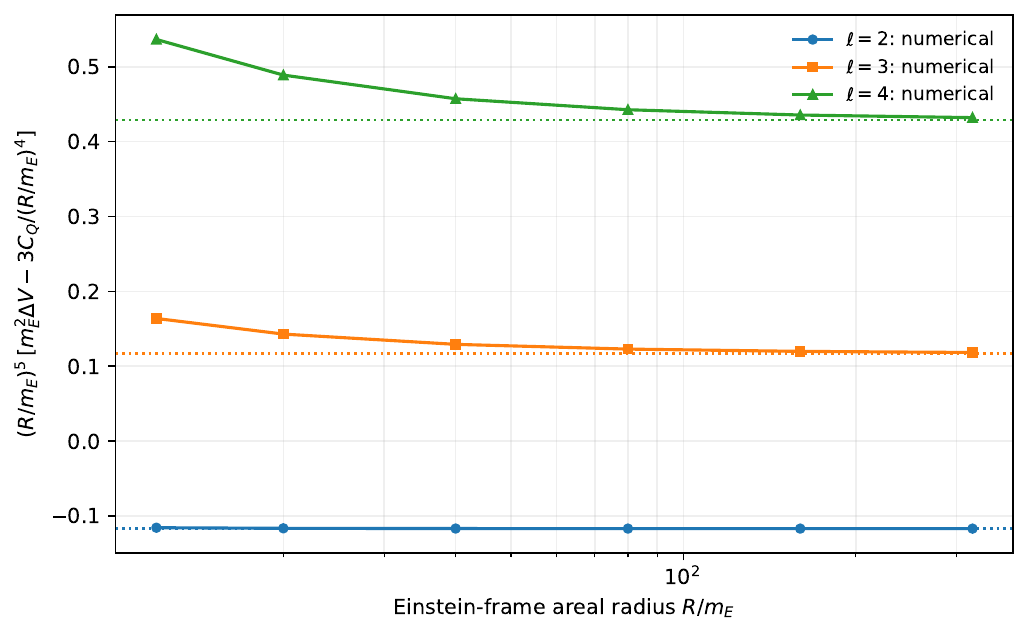}
\caption{Multipole-dependent next-order axial potential at fixed Einstein ADM mass ($Q=0.3$). Solid: exact-minus-leading correction multiplied by $(R/m_E)^5$; dotted: analytic limits $(L-9)C_Q/3$ from Eq.~\eqref{eq:axialnextorder}.}\label{fig:axialnext}
\end{figure}

Matching the asymptotic scalar charge in equation~\eqref{eq:chiweak} to \eqref{eq:jnwscalar} gives, for the weakly self-gravitating matter lens of equation~\eqref{eq:charge},
\begin{equation}
 m_E=\frac{1+6Q^2}{1+8Q^2}\,r_g,
\qquad \zeta^2=\frac{1+6Q^2}{1+8Q^2}.
\label{eq:jnwmatching}
\end{equation}
These are exterior asymptotic matching relations and do not specify the lens interior or hold for an object with a different effective scalar charge. The relation to the Jordan-frame post-Newtonian light-deflection parameter is particularly simple:
\begin{equation}
 s\equiv\zeta^2=\frac{1+\gamma_{\rm PPN}}{2},\qquad
 C_Q\equiv s^{-1}-1=\frac{1-\gamma_{\rm PPN}}{1+\gamma_{\rm PPN}}.
 \label{eq:matchedgamma}
\end{equation}
This identity follows algebraically from the locally normalized coupling in Eq.~\eqref{eq:gammaPPN}; it is specific to the unscreened weak-field matter source considered here. In particular, a comparison at fixed Einstein-frame mass $m_E$ is not interchangeable with one at fixed locally measured matter mass $M_L$. Expand the axial potential at fixed $m_E$ and fixed areal radius $R$ (rather than comparing two different radial coordinates). One obtains
\begin{align}
V_\ell^{\rm odd}(R)={}&\frac{L}{R^2}-2(L+3)\frac{m_E}{R^3}\notag\\
&+\left(9+\frac3{\zeta^2}\right)\frac{m_E^2}{R^4}
+O(m_E^3/R^5),\\
\Delta V_\ell^{\rm odd}(R)={}&
V_\ell^{\rm odd}-V_\ell^{\rm Schw}\notag\\
={}&\frac{6Q^2}{1+6Q^2}\frac{m_E^2}{R^4}
+O(m_E^3/R^5).
\label{eq:odd_delta}
\end{align}
Here and below $L=\ell(\ell+1)$. At the next radial order, the first explicit multipole dependence of the scalar-charge difference appears:
\begin{equation}
\begin{aligned}
\Delta V_{\ell}^{\rm odd}(R)\big|_{m_E}
={}&3C_Q\frac{m_E^2}{R^4}\\
&+\frac{L-9}{3}C_Q\frac{m_E^3}{R^5}
+O\!\left(\frac{m_E^4}{R^6}\right).
\end{aligned}
\label{eq:axialnextorder}
\end{equation}
This asymptotic coefficient is obtained by reverting the exact relation $R=r f^{(1-\zeta)/2}$ before expanding the potential; expanding directly at equal coordinate $r$ would give a different coefficient. The $R^{-5}$ term is negative for $\ell=2$ and positive for $\ell\ge3$ when $Q\ne0$, whereas the displayed $R^{-4}$ term is multipole independent. This relative sign change gives a direct benchmark for independent numerical implementations of the axial equation. Figure~\ref{fig:axialnext} compares the exact exterior potentials with the predicted next-order limits using high-precision root finding for $R(r)$. At $Q=0.3$ and $R=20m_E$, adding the $R^{-5}$ term reduces the relative asymptotic error from $1.69\%$ to $0.006\%$ for $\ell=2$ and from $2.00\%$ to $0.364\%$ for $\ell=3$. These are convergence checks of the exterior perturbation potential, rather than estimates of a detected signal.

Thus a parity-decoupled physical axial propagation potential contains an $O(Q^2)$ correction relative to the Schwarzschild reference at fixed Einstein-frame ADM mass. This is not a claimed $O(Q^2)$ observed lens-induced phase: physical two-image phases require a mapping from axial partial waves to a lensed wavepacket, the even-parity sector, consistent Jordan-frame source/detector conventions, and GR baseline subtraction with fixed observables. The result also relies on an exterior spherical ray, whereas a cluster-scale lens is generally extended and nonspherical.

\subsection{Mass normalization and a model-specific consistency relation}\label{sec:massmatching}
The axial potential difference in Eq.~\eqref{eq:odd_delta} can be expressed entirely in the weak-field PPN parameter when the Einstein-frame ADM mass is held fixed:
\begin{equation}
\Delta V_{\ell}^{\rm odd}(R)\big|_{m_E}
=3\frac{1-\gamma_{\rm PPN}}{1+\gamma_{\rm PPN}}
\frac{m_E^2}{R^4}+O(m_E^3/R^5).
\label{eq:oddppnrelation}
\end{equation}
This relation is independent of the multipole $\ell$ at the displayed radial order; the full potential itself remains $\ell$ dependent. The parameter relation is a consistency check between the weak-field matter coupling and the strong-background exterior perturbation calculation, not a new independent measurement of $\gamma_{\rm PPN}$.

For comparison, if the same potential is re-expanded at fixed $r_g=G_NM_L/c^2$ and the Schwarzschild reference is instead assigned mass $r_g$, the two expansions give
\begin{equation}
\begin{aligned}
\Delta V_\ell^{\rm odd}(R)\big|_{r_g}
={}&(L+3)(1-\gamma_{\rm PPN})\frac{r_g}{R^3}\\
&+(9s^2+3s-12)\frac{r_g^2}{R^4}\\
&+O(r_g^3/R^5).
\end{aligned}
\label{eq:oddGNmatched}
\end{equation}
with $s=(1+\gamma_{\rm PPN})/2$. The leading $R^{-3}$ term is a consequence of using different Einstein-frame masses for a fixed Jordan-frame matter mass. Equation~\eqref{eq:oddGNmatched} is a mass-convention diagnostic on a common Einstein-frame areal-radius coordinate, not an observed Jordan-frame lens comparison: a physical equal-mass test also requires a common operational radius and source/observer calibration. The two formulas must therefore not be combined as if they described the same subtraction.

\begin{figure}[!htbp]\centering
\includegraphics[width=.90\columnwidth]{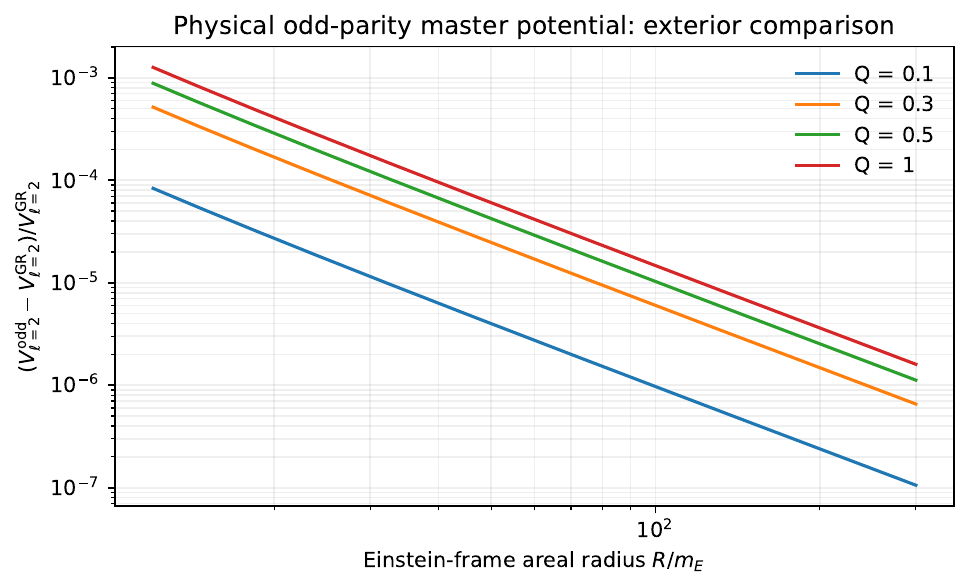}
\caption{Relative $\ell=2$ axial-potential difference from Schwarzschild at fixed ADM mass and circumferential radius. Exact curves follow Eq.~\eqref{eq:oddexplicit}; the weak-exterior form is Eq.~\eqref{eq:odd_delta}.}\label{fig:axial}
\end{figure}

Figure~\ref{fig:axial} illustrates the equal-$m_E$ potential difference evaluated from the exact exterior solution. The weak-field coefficients in Eqs.~\eqref{eq:oddppnrelation} and \eqref{eq:oddGNmatched} are independently checked by symbolic expansion in the accompanying verification code.

\subsection{Finite-radius WKB phase of the axial mode}\label{sec:axialphase}
The local potential difference in equation~\eqref{eq:odd_delta} is not by itself a phase. We can, however, take one controlled step further within the exact exterior odd-parity master equation: define the finite-interval radial WKB action at fixed multipole $\ell$, fixed asymptotically normalized Einstein-frame frequency $\omega$, and fixed Einstein-frame areal radii $R_a<R_b$,
\begin{equation}
\begin{aligned}
\Phi_\ell^{\rm WKB}(\omega;R_a,R_b;\zeta)
={}&\int_{R_a}^{R_b}\!J_\zeta(R)
\sqrt{\omega^2-V_\ell^{\rm odd}}\,\dd R,\\
J_\zeta(R)&=\dd r_*/\dd R.
\end{aligned}
\label{eq:radialaction}
\end{equation}
The square root is taken on the oscillatory branch; we require $\omega^2>\max V_\ell^{\rm odd}$ throughout the interval, $\omega R_a\gg\ell$, and $R_a$ outside any matter surface. In the FJNW exterior,
\begin{equation}
 J_\zeta(R)=\frac{f(r)^{-(1+\zeta)/2}}
 {1+(1-\zeta)B/[2rf(r)]},\qquad R=r f(r)^{(1-\zeta)/2}.
 \label{eq:jacobianaxial}
\end{equation}
Equation~\eqref{eq:radialaction} is a radial partial-wave WKB diagnostic, not the asymptotic scattering phase (which also requires boundary conditions) and not the observed difference between two gravitationally lensed images. It is nevertheless defined from the physical odd-parity operator, rather than from the longitudinal scalar surrogate.

At fixed $\ell$ the large-$\omega$ difference relative to equal-mass Schwarzschild data on the same areal-radius interval has the expansion
\begin{align}
\Delta\Phi_\ell^{\rm WKB}
={}&\omega\int_{R_a}^{R_b}\!\Delta J\,\dd R\notag\\
&-\frac{1}{2\omega}\int_{R_a}^{R_b}\!\Delta(JV_\ell)\,\dd R
+O(\omega^{-3}),\label{eq:axialphaseexp}\\
\Delta J={}&J_\zeta-J_1,\notag\\
\Delta(JV_\ell)={}&J_\zeta V_{\ell,\zeta}^{\rm odd}
-J_1 V_{\ell,1}^{\rm odd}.\nonumber
\end{align}
The first term is proportional to frequency and belongs to the achromatic optical-path part. The $\omega^{-1}$ term is the first formal frequency-dependent contribution to this fixed-$\ell$ radial action. Crucially, the potential must be multiplied by the optical Jacobian before comparing backgrounds: subtracting $V_{\ell,\zeta}-V_{\ell,1}$ alone gives a different answer.

Let $C_Q\equiv\zeta^{-2}-1=2Q^2/(1+6Q^2)$, $m\equiv m_E$, and $L\equiv\ell(\ell+1)$. Expanding at fixed $m$ and areal radius $R\gg m$ gives
\begin{align}
\Delta J={}&-\frac{C_Qm^2}{2R^2}
-\frac{8C_Qm^3}{3R^3}+O(m^4/R^4),
\label{eq:jacexp}\\
\Delta(JV_\ell)={}&-\frac{C_Q(L-6)m^2}{2R^4}\notag\\
&-\frac{2C_Q(2L-9)m^3}{3R^5}
+O(m^4/R^6).
\label{eq:momentexp}
\end{align}
Here $J_1V_{\ell,1}^{\rm odd}=L/R^2-6m/R^3$ exactly. Combining equations~\eqref{eq:axialphaseexp}--\eqref{eq:momentexp}, with $\Delta_n\equiv R_a^{-n}-R_b^{-n}$, yields
\begin{equation}
\begin{aligned}
\Delta\Phi^{\rm WKB}_\ell={}&-\frac{C_Q\omega m^2}{2}\Delta_1
+\frac{C_Q(L-6)m^2}{12\omega}\Delta_3\\
&+O(\omega m^3/R_a^2)
+O(m^3/(\omega R_a^4))\\
&+O(\omega^{-3}).
\end{aligned}
\label{eq:axialwkbmain}
\end{equation}
For the lowest gravitational multipole $\ell=2$ ($L=6$), the displayed $m^2/\omega$ term cancels even though the local $\Delta V_\ell$ in equation~\eqref{eq:odd_delta} is nonzero. The next exterior term in the same fixed-$\ell$ WKB potential moment gives
\begin{equation}
\left.\Delta\Phi_{\ell,\,\omega^{-1}}^{\rm WKB}\right|_{\ell=2}
 =\frac{C_Q m^3}{4\omega}\Delta_4
 +O(m^4/(\omega R_a^5))+O(\omega^{-3}).
\label{eq:axialquadrupole}
\end{equation}
This is a cancellation in the radial WKB action under the stated matching convention; it is neither a universal cancellation of gravitational-wave dispersion nor a prediction of zero image-differential phase. The expansion $\omega\to\infty$ was taken at fixed $\ell$; it must not be interchanged without justification with the high-$\ell$ limit $\ell\sim\omega b$ relevant for a finite-impact-parameter lensing wavepacket. Any source/observer master-field conversion, interior matching, even-parity response and endpoint phases must also be specified before extracting an observable.

\begin{figure}[!htbp]\centering
\includegraphics[width=.90\columnwidth]{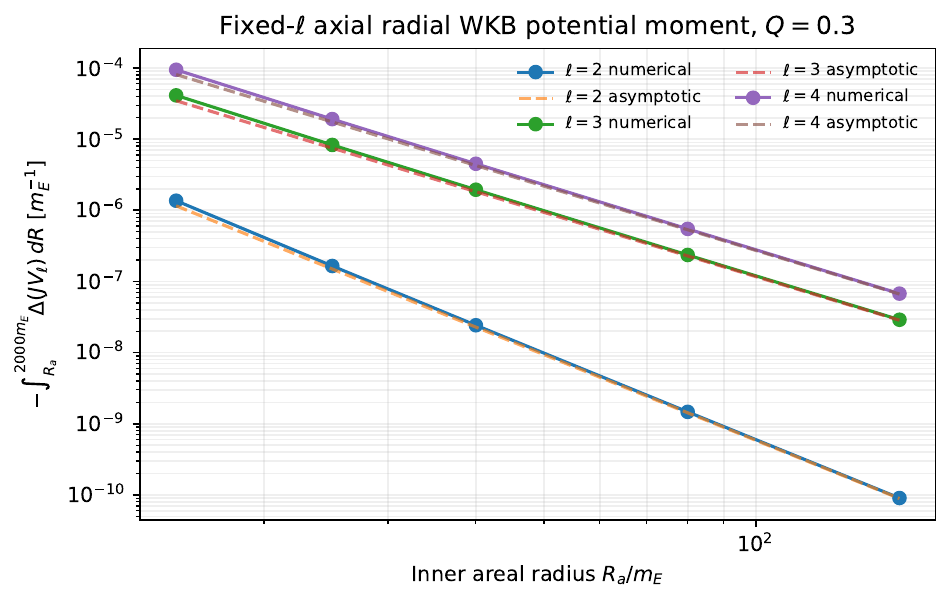}
\caption{Jacobian-weighted axial potential moment for $R\in[R_a,2000m_E]$ ($Q=0.3$). Solid: exact exterior; dashed: leading weak-exterior term. The quadrupole starts at a higher radial order than $\ell\ge3$.}\label{fig:axialmom}
\end{figure}

Figure~\ref{fig:axialmom} compares the exact radial potential moments with their exterior expansions. For an independent check we evaluate equation~\eqref{eq:radialaction} through its local momentum, using the exact FJNW $R(r)$ relation and $V_\ell$, and separately integrate the two moments in equation~\eqref{eq:axialphaseexp}. At $Q=0.3$, $R_a=80m$, $R_b=2000m$, the numerical weighted-potential moments in units of $m^{-1}$ are $-1.4700\times10^{-9}$ for $\ell=2$ and $-2.3558\times10^{-7}$ for $\ell=3$. The distinct radial scaling and the $\ell=2$ cancellation agree with equation~\eqref{eq:momentexp}. The finite-$\omega$ local-momentum quadrature at $\omega m=5$ agrees with the leading $\omega^{-1}$ term to within the explicitly tabulated remainder in the accompanying code.

\paragraph{Fixed-impact-parameter eikonal limit and bending.}\label{sec:fixedimpact}
The preceding fixed-$\ell$ limit is not the impact-parameter limit of a lensed wavepacket. For a large-multipole sequence, set
\begin{equation}
\begin{aligned}
b&=\sqrt{\ell(\ell+1)}/\omega,\\
H_\zeta(R)&=A_\zeta(R)/R^2,\\
W_\zeta(R)&=-6A_\zeta(R)m_\zeta(R)/R^3,\\
A_\zeta&=f^\zeta.
\end{aligned}
\label{eq:fixedbdef}
\end{equation}
The exact axial potential is then $V_\ell^{\rm odd}=\omega^2b^2H_\zeta+W_\zeta$. On an exterior finite annulus $R_a<R<R_b$ obeying $K_\zeta(R,b)\equiv1-b^2H_\zeta(R)>\kappa>0$ (and the analogous inequality in the Schwarzschild reference), the same physical axial equation gives the controlled fixed-$b$ expansion
\begin{align}
\Phi_{\rm ann}^{\rm WKB}={}&\omega\int_{R_a}^{R_b}
J_\zeta\sqrt{K_\zeta}\,\dd R\notag\\
&-\frac{1}{2\omega}\int_{R_a}^{R_b}
\frac{J_\zeta W_\zeta}{\sqrt{K_\zeta}}\,\dd R
+O(\omega^{-3}),\label{eq:fixedbaction}\\
\Delta\Phi_{\rm ann}^{\rm WKB}={}&\omega\Delta I_{\rm geo}(b)\notag\\
&-\frac{\Delta I_{\rm odd}(b)}{2\omega}
+O(\omega^{-3}).
\label{eq:fixedbdelta}
\end{align}
where the $I$'s denote precisely the first and second radial integrals in equation~\eqref{eq:fixedbaction} and $\Delta$ subtracts the equal-$m_E$ Schwarzschild annulus. The first integral is the geometrical optical path, while the second is an axial-master-field $1/\omega$ correction on a turning-point-free segment. This is not an asymptotic lens transfer function: the expansion of $K^{-1/2}$ is nonuniform at a closest-approach turning point, and an Airy/uniform matching construction plus source and receiver data would be necessary. In particular, the $\ell=2$ cancellation in equation~\eqref{eq:axialquadrupole} cannot be carried over to this distinct $\ell\sim\omega b$ limit.

To check the leading optical term against a full ray geometry, consider an equatorial null geodesic in the exact FJNW exterior. Conformal rescaling preserves null curves, and the asymptotic normalization $A(\chi_\infty)=1$ allows the asymptotic Einstein-frame impact parameter to be used here. If $r_0$ is the exterior turning radius, the conserved energy and angular momentum imply
\begin{equation}
\begin{aligned}
b&=r_0 f(r_0)^{(1-2\zeta)/2},\\
\widehat\alpha(b;\zeta)+\pi
&=2\int_{r_0}^\infty
\frac{b\,\dd r}{r^2f^{1-\zeta}
\sqrt{1-b^2f^{2\zeta-1}/r^2}}.
\end{aligned}
\label{eq:jnwbendingexact}
\end{equation}
This null-geodesic result is an achromatic geometrical-optics property, not a spin-two finite-frequency dispersive phase; the FJNW geometry and its lensing properties are established in the literature \citep{Perlick2004,ChowdhuryBanerjee2020}. Nevertheless, it permits a useful exact-to-asymptotic check of the earlier point-lens interpretation.

For $x=m_E/b\ll1$, set $\epsilon=B/r_0$ and $u=r_0/r$. Then equation~\eqref{eq:jnwbendingexact} becomes
\begin{equation}
 \widehat\alpha+\pi=2\int_0^1\frac{(1-\epsilon)^{(1-2\zeta)/2}(1-\epsilon u)^{\zeta-1}\,\dd u}
 {\sqrt{1-u^2[(1-\epsilon u)/(1-\epsilon)]^{2\zeta-1}}}.
 \label{eq:bendexpandedintegral}
\end{equation}
Expanding this integrand in $\epsilon$ and using the elementary integrals
\begin{align*}
\int_0^1\frac{\dd u}{\sqrt{1-u^2}(1+u)}&=1,\\
\int_0^1\frac{\dd u}{\sqrt{1-u^2}(1+u)^2}&=\frac23,
\end{align*}
yields
\begin{align}
 \widehat\alpha&=2\zeta\epsilon+
 \bigl(\pi\zeta^2-\pi/16-2\zeta^2+\zeta\bigr)\epsilon^2+O(\epsilon^3),\\
 \epsilon&=\frac{2x}{\zeta}+\frac{4\zeta-2}{\zeta^2}x^2+O(x^3).
\end{align}
Therefore, at fixed Einstein-frame ADM mass $m_E$ and asymptotic $b$,
\begin{align}
\widehat\alpha(b;\zeta)={}&\frac{4m_E}{b}\notag\\
&+\frac{\pi}{4}(16-\zeta^{-2})\frac{m_E^2}{b^2}
+O(m_E^3/b^3),\label{eq:fixedb2pn}\\
\widehat\alpha(b;Q)-\widehat\alpha(b;0)
={}&-\frac{\pi Q^2}{2(1+6Q^2)}\frac{m_E^2}{b^2}\notag\\
&+O(m_E^3/b^3).\label{eq:fixedb2pndelta}
\end{align}
The leading $4m_E/b$ term is equal to GR in this Einstein-ADM-mass-matched comparison; scalar charge first enters this bending comparison at second post-Minkowskian order, not at a frequency-dependent $1/\omega$ order. In the distinct, physically measured Jordan-frame mass convention $m_E=[(1+6Q^2)/(1+8Q^2)]r_g$, the leading term becomes $2(1+\gamma_{\rm PPN})r_g/b$, agreeing with equation~\eqref{eq:pnalpha}. One must not mix these two mass-matching conventions when declaring a leading-order correction. In particular, neither equation~\eqref{eq:fixedb2pndelta} nor the annular coefficient in equation~\eqref{eq:fixedbdelta} is an observed two-image GW chromatic response.

\begin{figure}[!htbp]\centering
\includegraphics[width=.90\columnwidth]{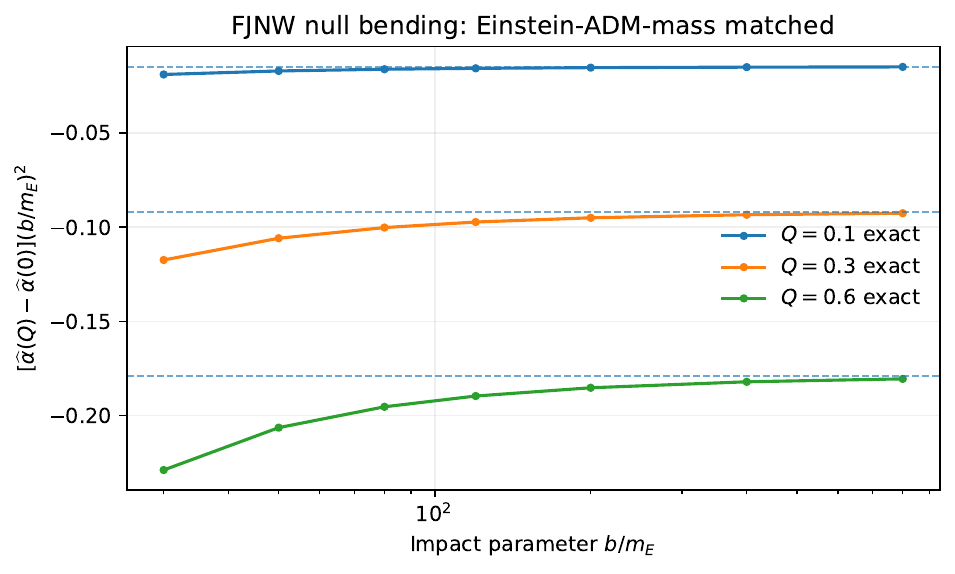}
\caption{FJNW-minus-Schwarzschild null-geodesic bending, scaled by $(b/m_E)^2$, for equal Einstein ADM mass and impact parameter. Dashed: analytic asymptote $-\pi(\zeta^{-2}-1)/4$.}\label{fig:fixedb}
\end{figure}

Direct integration of equation~\eqref{eq:jnwbendingexact} at $Q=0.1,0.3,0.6$ independently approaches the coefficient $-\pi(\zeta^{-2}-1)/4$ as $b/m_E$ increases; figure~\ref{fig:fixedb} gives the GR-subtracted, impact-rescaled comparison. The same audit independently tests equation~\eqref{eq:fixedbdelta} over $\ell=20,40,80,160$, two impact parameters and three couplings in exterior, turning-point-free intervals. These tests show that the optical-subtracted axial WKB residual scales as $1/\omega$ within the defined annulus; they neither perform a partial-wave sum nor furnish a two-image likelihood.

\subsection{Comparison with a scalar surrogate}\label{sec:surrogate}\label{sec:3d}\label{sec:ppn}
For comparison with the physical odd-parity master equation~\eqref{eq:oddmaster}, consider the following longitudinal flux-weighted scalar equation:
\begin{equation}
\begin{aligned}
h''+(\ln F)'h'+k_L^2h&=0,\\
F=M_*^2,\quad k_L&=2\pi(1+z_L)f/c.
\end{aligned}\label{eq:oneD}
\end{equation}
Here a prime means a derivative along the straight ray, and $z_L$ is the lens redshift. The rescaling $u=\sqrt{F}h$ yields a formal potential $V_F=\tfrac12(\ln F)''+\tfrac14[(\ln F)']^2$. In first-order WKB, with vanishing derivative boundary term, $\delta\Psi_{1\rm D}=-[8k_L]^{-1}\int[(\ln F)']^2\,\dd z$. For $r^2=b^2+z^2$ and equation~\eqref{eq:charge},
\begin{equation}
\begin{aligned}
\delta\Psi_{1\rm D}(f,b)
&=-\frac{c r_g^2}{8(1+z_L)fb^3}\kappa_Q,\\
\kappa_Q&=\frac{Q^4}{(1+8Q^2)^2},\quad
\delta\Psi=\beta(f_0/f).
\end{aligned}
\label{eq:phase}
\end{equation}
Here $f_0$ is a chosen reference frequency and $\beta$ a dimensionless phenomenological phase coefficient. These results assume longitudinal dynamics and cannot be promoted to a Horndeski tensor prediction. Indeed, applying the same $F(\boldsymbol x)$ to a distinct three-dimensional flat-space scalar action leads after the rescaling $u=\sqrt Fh$ to
\begin{equation}
 V_{3\rm D}=\tfrac12\nabla^2\ln F+\tfrac14|\nabla\ln F|^2,
 \qquad \delta\Psi_{3\rm D}^{\rm sc}=4\delta\Psi_{1\rm D},\label{eq:v3d}
\end{equation}
for the exterior $1/r$ profile and $b>0$. The factor four is a counterexample to surrogate robustness, not a physical theory error bar. The standard weak-field massless scalar--tensor result \citep{Clifton2012,Will2014}, with the same locally normalized coupling, is
\begin{equation}
 \gamma_{\rm PPN}=\frac{1+4Q^2}{1+8Q^2},\qquad
 \kappa_Q=\frac{(1-\gamma_{\rm PPN})^2}{16}.\label{eq:gammaPPN}
\end{equation}
For the illustrative scale estimate we impose $|\gamma_{\rm PPN}-1|\le10^{-4}$, not a new Cassini confidence bound. The Cassini measurement \citep{Bertotti2003} and this algebraic identity provide a conditional surrogate scale, not an observational limit on a full tensor coupling.

\begin{figure*}[!t]\centering
\includegraphics[width=.92\textwidth]{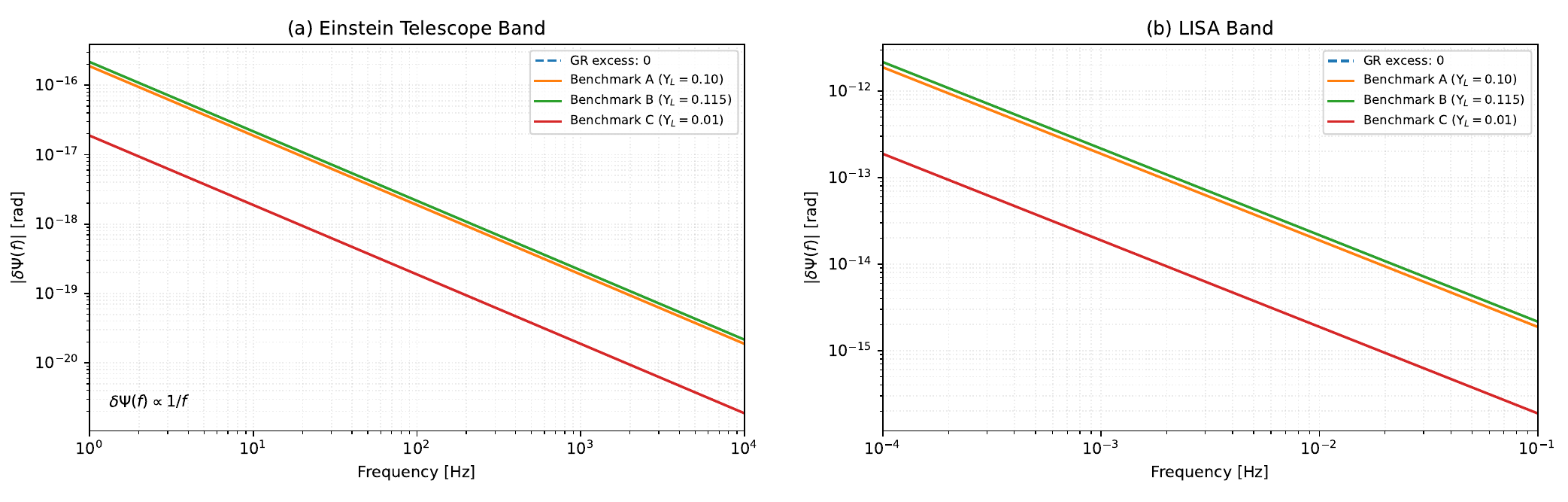}
\caption{Illustrative $1/f$ phase curves in the ET (a) and LISA (b) frequency bands. The labels $\Upsilon_L=0.10,0.115,0.01$ specify illustrative surrogate inputs, not inferred values of the coupling $Q$. The zero GR-excess reference is absent from the logarithmic vertical scale.}\label{fig:historicalphase}
\end{figure*}

Figure~\ref{fig:historicalphase} shows only the frequency dependence of the assumed scalar-surrogate phase. The plotting symbol $\Upsilon_L$ labels the benchmark curves and is not the action coupling $Q$ or a measured $\Delta\beta$; the latter is introduced as an independent parameter in section~\ref{sec:two}. We now turn from this conditional phase illustration to the achromatic lens geometry that follows directly from the specified weak-field matter coupling.

\section{Weak-field lens geometry and mass identifiability}\label{sec:pnlens}
The lens geometry used to compare two paths is itself a physical prediction of the specified unscreened model. To first post-Newtonian order around a weakly self-gravitating, static, approximately isolated mass, write the Jordan-frame metric in asymptotically isotropic coordinates as
\begin{equation}
\begin{aligned}
\dd s_J^2={}&-(1-2u)c^2\dd t^2
+(1+2\gamma_{\rm PPN}u)\dd\bm x^2\\
&+O(u^2),\quad u=G_NM_L/(c^2r).
\end{aligned}
\label{eq:pnmetric}
\end{equation}
Null rays in this metric have refractive index $n=1+(1+\gamma_{\rm PPN})u+O(u^2)$, yielding the physical small-angle deflection
\begin{equation}
\hat\alpha(b)=\frac{2(1+\gamma_{\rm PPN})G_NM_L}{c^2b}+O(u^2).
\label{eq:pnalpha}
\end{equation}
For an ideal thin point lens with angular-diameter distances $D_{\rm OL},D_{\rm OS},D_{\rm LS}$, the corresponding Einstein radius obeys
\begin{equation}
\begin{aligned}
(R_E^{\rm PN})^2
&=\frac{2(1+\gamma_{\rm PPN})G_NM_L D_{\rm OL}D_{\rm LS}}{c^2D_{\rm OS}},\\
\frac{R_E^{\rm PN}}{R_E^{\rm GR}}
&=\sqrt{(1+\gamma_{\rm PPN})/2}.
\end{aligned}
\label{eq:pnRE}
\end{equation}
At fixed dimensionless source position $y\equiv\beta_{\rm src}/\theta_E^{\rm PN}$ the image positions $x_\pm=(y\pm\sqrt{y^2+4})/2$ and signed magnifications take their usual point-lens forms. Define the positive delay-shape function $\mathcal T(y)\equiv y\sqrt{y^2+4}/2+\ln[(\sqrt{y^2+4}+y)/(\sqrt{y^2+4}-y)]$. The achromatic image arrival-time difference, derived from the same Fermat potential, is
\begin{equation}
\begin{aligned}
\Delta t_{\rm PN}&=\frac{2(1+\gamma_{\rm PPN})G_NM_L}{c^3}
(1+z_L)\mathcal T(y),\\
\frac{\Delta t_{\rm PN}}{\Delta t_{\rm GR}}
&=(1+\gamma_{\rm PPN})/2.
\end{aligned}
\label{eq:pndelay}
\end{equation}
The formulas hold at first post-Newtonian order in a massless unscreened model with weakly self-gravitating lens matter; they are not a full dynamical scalar--tensor lens solution. At fixed $y$ and lens mass, the relative time-delay shift is $(\gamma_{\rm PPN}-1)/2$, whereas the relative Einstein-radius shift is $\sqrt{(1+\gamma_{\rm PPN})/2}-1$. Under the illustrative condition $|\gamma_{\rm PPN}-1|\le10^{-4}$, the delay shift is no larger than $5\times10^{-5}$ and the Einstein-radius shift no larger than approximately $2.51\times10^{-5}$ in magnitude. This is an action-linked, leading $O(Q^2)$ achromatic lensing channel, distinct from the $O(Q^4)$ corrections predicted by the scalar phase surrogates. It is not independently informative about $Q$ without an external calibration of lens mass, profile and distances.

For consistency, if the same weak-field surrogate phase expression is evaluated at the PN-corrected impact parameters $b_\pm=R_E^{\rm PN}|x_\pm|$ instead of the GR reference parameters, its coefficient changes as
\begin{equation}
\begin{aligned}
C_d^{\rm PN}(M_L,\gamma_{\rm PPN})
&=C_d^{\rm GR}(M_L)
\left(\frac{2}{1+\gamma_{\rm PPN}}\right)^{3/2},\\
&\hspace{9mm} d\in\{1\mathrm D,3\mathrm D\}.
\end{aligned}
\label{eq:pnC}
\end{equation}
This is only a geometric consistency rescaling of the surrogate, not a recovered Horndeski propagation kernel. In the near-GR limit the correction to the surrogate's leading $Q^4$ term is $O(Q^6)$. It is inappropriate, however, to use GR impact parameters together with an unrestricted finite-$Q$ envelope without acknowledging this geometrical difference.

\subsection{Achromatic mass--gravity degeneracy}\label{sec:massdeg}
The first-PN rescaling in section~\ref{sec:pnlens} does not imply that the lens geometry alone measures the fundamental scalar coupling. In a point-lens model with known angular-diameter distances and a constant weak-field $\gamma_{\rm PPN}$, introduce the effective lensing mass
\begin{equation}
\begin{aligned}
{\cal M}_{\rm lens}
&=M_L(1+\gamma_{\rm PPN})/2,\\
\theta_E^2&=\frac{4G_N{\cal M}_{\rm lens}D_{\rm LS}}
{c^2 D_{\rm OL}D_{\rm OS}},\\
\Delta t&=\frac{4G_N{\cal M}_{\rm lens}}{c^3}
(1+z_L){\cal T}(y).
\end{aligned}
\label{eq:effmass}
\end{equation}
The usual point-lens equation is $\beta_{\rm src}=\theta-\theta_E^2/\theta$; its image positions and signed magnifications depend on $\theta_E$ and the source angle $\beta_{\rm src}$, hence on $\mathcal M_{\rm lens}$ rather than separately on $(M_L,\gamma_{\rm PPN})$. The same is true of the image-arrival delay because both the gravitational and geometric pieces enter through the same Fermat potential. This statement holds whether the unknown source angle or the dimensionless $y=\beta_{\rm src}/\theta_E$ is used as a nuisance parameter: one can vary $M_L$ and $\gamma_{\rm PPN}$ while holding $\mathcal M_{\rm lens}$ and $\beta_{\rm src}$ fixed.

\noindent\textbf{Rank-deficiency statement.} Within the stated first-PN, static, isolated point-lens model, and without an independent dynamical mass estimate, the collection of achromatic image angles, magnification ratios and time delays cannot identify $M_L$ and $\gamma_{\rm PPN}$ separately. To see this algebraically, set $m=\ln M_L$, $g=\gamma_{\rm PPN}$ and $\lambda_m=\ln\mathcal M_{\rm lens}=m+\ln[(1+g)/2]$. Any lens-only likelihood of the form ${\cal L}_{\rm lens}(\lambda_m,\beta_{\rm src},\ldots)$ depends on $(m,g)$ only through $\lambda_m$. Conditional on fixed other parameters, if $I_{\lambda_m}$ denotes its Fisher information for $\lambda_m$, the corresponding block is
\begin{equation}
\begin{gathered}
\Gamma_{(m,g)}\propto
\begin{pmatrix}
1&(1+g)^{-1}\\
(1+g)^{-1}&(1+g)^{-2}
\end{pmatrix},\\
\det\Gamma_{(m,g)}=0,\qquad
\delta m=-\frac{\delta g}{1+g}.
\end{gathered}
\label{eq:massfisher}
\end{equation}
The last equality gives its exact local null direction. Additional image-angle or time-delay measurements of the same ideal point lens can improve the estimate of $\lambda_m$, but cannot turn this block into a full-rank two-parameter measurement. In particular, the point-lens ratio
\begin{equation}
 \frac{\Delta t}{R_E^2}=\frac{1+z_L}{c}\frac{D_{\rm OS}}{D_{\rm OL}D_{\rm LS}}{\cal T}(y)
 \label{eq:masscancel}
\end{equation}
is independent of $M_L$ and $\gamma_{\rm PPN}$ once distances and $y$ are specified. This mass--potential degeneracy is related to the well-known need for independent stellar kinematics or other mass information in strong-lens tests of the PPN parameter \citep{Guo2026LensingPPN}. Equation~\eqref{eq:massfisher} specializes the established mass--potential degeneracy to the present point-lens identifiability experiment.

\paragraph{External mass information.}\label{sec:massprior}
Suppose a lensing analysis delivers an approximately Gaussian independent estimate of $\lambda_m$ with standard deviation $s_\lambda$, and an external dynamical or other non-lensing measurement delivers an independent Gaussian prior on $m$ with standard deviation $s_m$. Linear error propagation, or inversion of the two-measurement Fisher matrix, then yields
\begin{equation}
\begin{aligned}
\sigma_g&=(1+g)\sqrt{s_\lambda^2+s_m^2},\\
\left.\sigma_g\right|_{g=1}
&=2\sqrt{s_\lambda^2+s_m^2}\geq2s_m.
\end{aligned}
 \label{eq:massprior}
\end{equation}
The uncertainties here are fractional logarithmic mass errors and a local Gaussian approximation to $g$; they are not the result of fitting an observed lens. For illustration, if $s_m=10^{-2}$, $10^{-3}$ or $10^{-4}$, the ideal $s_\lambda\to0$ floors are, respectively, $\sigma_g=2\times10^{-2}$, $2\times10^{-3}$ or $2\times10^{-4}$. A statistical target $\sigma_g\lesssim10^{-5}$ around GR would require $s_m\lesssim5\times10^{-6}$ even if the achromatic lens strength were measured perfectly. These idealized floors make no claim about achievable stellar-kinematic accuracies or extragalactic constraints; realistic extended lenses introduce further model and distance uncertainties. Existing extra-galactic analyses break the degeneracy by combining lensing with dynamics or additional distance information \citep{Guo2026LensingPPN}.

In the same specified conformal theory, $Q^2=(1-g)/(8g-4)$ with $g\in(1/2,1]$. Near $Q=0$, $1-g=4Q^2+O(Q^4)$ and hence $\partial g/\partial Q=0$. A regular Gaussian conversion $\sigma_Q=\sigma_g/|\partial g/\partial Q|$ therefore fails at GR. If one includes a validated chromatic phase prediction in the future, its additional parameter dependence could in principle break the achromatic mass degeneracy, but the scalar surrogates in section~\ref{sec:3d} are not such a validated signal. The existing $\Delta\beta$ Fisher calculation cannot, by itself, be converted into a bound on $Q$.

\subsection{Source-branch consistency}\label{sec:sourcebranch}
The source-generation question cannot be settled by assigning an arbitrary $f^{-1}$ coefficient to the binary. For the same conformal action, the exterior Einstein-frame equations~\eqref{eq:backgroundE} admit the exact branch
\begin{equation}
\begin{aligned}
\chi=\chi_\infty&=\mathrm{constant},\quad R^E_{\mu\nu}=0,\\
g^J_{\mu\nu}&=A^2(\chi_\infty)g^E_{\mu\nu}.
\end{aligned}
 \label{eq:grbranch}
\end{equation}
Consequently every vacuum GR solution is also a solution on this constant-scalar branch, up to the constant metric rescaling. A vacuum black-hole binary selected on this branch, with no incoming scalar radiation and constant asymptotic scalar, has the same vacuum binary evolution and tensor waveform dynamics as its GR counterpart; the action does not require a new scalar-generation $f^{-1}$ phase for that source. This known GR branch is used here as a source-model consistency check, not claimed as a new no-hair theorem. It is an existence statement about a specified solution branch, not a theorem that all black-hole binaries in arbitrary scalar environments coincide with GR. Cosmological scalar drift, nontrivial initial scalar data and matter environments require separate treatment \citep{Will2014,Clifton2012}.

By contrast, the lens considered below is a weakly self-gravitating matter distribution. Variation of the matter action $S_m[A^2(\chi)g^E,\psi_m]$ gives schematically in the lens interior
\begin{equation}
 \Box_E\chi=-\frac{\alpha(\chi)}{\mpl}T_E,
 \qquad \alpha(\chi)=\mpl\frac{\dd\ln A}{\dd\chi},
 \label{eq:mattertrace}
\end{equation}
so matter can source a quasistatic scalar profile even though a distant vacuum binary is on the constant-scalar branch. Here $T_E$ is the Einstein-frame matter stress-tensor trace, with conventions fixed by this variation. The point-mass lens is therefore a thin-lens idealization of ordinary matter, not a claim that an isolated black hole of the same mass carries the scalar charge in equation~\eqref{eq:charge}. For a distant lens, it can be useful to approximate the black-hole generation region by the constant-scalar vacuum branch while retaining a matter-sourced scalar profile near the lens. This is a scale-separation approximation, not an exact global solution with a strictly constant scalar in one open region and a nontrivial scalar everywhere else; the lens's physical GR-subtracted tensor transfer function remains uncomputed. For a resolved image pair, the physical quantity ultimately required is the frequency-dependent difference $\delta\varphi_{+-}(f)=\arg[T_+(f)/T_-(f)]-\arg[T^{\rm GR}_+(f)/T^{\rm GR}_-(f)]$, defined with a common source, observer and image-labeling prescription. No equation in the present spherical or reduced Fisher analysis evaluates this quantity.

This distinction changes how the following Fisher comparison should be interpreted. The unrestricted common source-phase coefficient $\gamma_{-1}$ is a phenomenological stress-test nuisance, not a prediction of equation~\eqref{eq:model} for the constant-scalar vacuum black-hole branch. On that branch it is consistent to set the additional source-generation coefficient to zero, so the advertised exact single-image source-versus-lens degeneracy need not arise. A two-image difference still offers a separately defined path comparison, but there is no demonstrated model-specific gain over an ordinary event. For neutron-star or mixed binaries, nontrivial matter sensitivities and source-generation waveforms must instead be derived before any coupling-level comparison.\par

The mass-normalized lens map thus determines the achromatic baseline, while the source-branch analysis shows why a same-power source phase cannot be assumed to occur in every binary. The next section isolates the statistical identifiability question without treating such a phenomenological phase as an action-derived signal.

\section{Two-image phase identifiability and Fisher diagnostics}\label{sec:two}
To test a potential degeneracy without assuming a specific binary-generation mechanism, introduce an unrestricted same-power common phase. As shown in section~\ref{sec:sourcebranch}, this is not compulsory for a vacuum black-hole source on the constant-scalar branch. The complex waveform is written as
\begin{equation}
\begin{aligned}
\tilde h_j(f)={}&F_j^{\rm GR}(f;\lambda_L)
T_{\rm com}(f)\tilde h_{\rm GR}(f;\theta_s)\\
&\times\exp\!\left[
 i(\gamma_{-1}+\beta_j)\frac{f_0}{f}\right].
\end{aligned}
\label{eq:wave}
\end{equation}
where $\gamma_{-1}$ describes a common source-side (or common-propagation) same-power phase, while $\beta_j$ denotes the excess phase along lens path $j$. Other source corrections are not claimed to vanish; the model isolates a possible exact same-power degeneracy. For one image, only $c_j=\gamma_{-1}+\beta_j$ can be measured, because $\partial_{\gamma_{-1}}\tilde h_j=\partial_{\beta_j}\tilde h_j=i(f_0/f)\tilde h_j$. The single-image Fisher matrix for these two unrestricted coefficients has rank one.

For two images define
\begin{equation}
c_{\rm com}=\gamma_{-1}+\tfrac12(\beta_++\beta_-),\qquad
\bta=\beta_+-\beta_-.
\end{equation}
In the local Gaussian-noise Fisher approximation \citep{Cutler1994,Vallisneri2008,Carson2020}, if noise from the two data segments is independent and the GR transfer functions are fitted, the shared phase factor cancels formally from the image ratio. With all other parameters held fixed, set $I_j=(if_0\tilde h_j/f\,|\,if_0\tilde h_j/f)>0$. Then
\begin{equation}
\begin{gathered}
\Gamma_{(c_{\rm com},\Delta\beta)}
=\begin{pmatrix}
I_++I_-&(I_+-I_-)/2\\
(I_+-I_-)/2&(I_++I_-)/4
\end{pmatrix},\\
\det\Gamma=I_+I_->0.
\end{gathered}
\label{eq:fisher_rank}
\end{equation}
and
\begin{equation}
\sigma_{\Delta\beta}^{2}=I_+^{-1}+I_-^{-1}.
\label{eq:variance}
\end{equation}
This is a quantitative rank/identifiability benefit, not evidence that a lensed pair offers a stronger coupling bound than an unlensed event. A full comparison must use a shared action-level generation and propagation model with intrinsic, lens and detector nuisance parameters. These are not marginalized in \eqref{eq:variance}. Importantly, observing both images requires an astrophysically feasible image delay.

\paragraph{Single-image versus two-image information.}\label{sec:directcompare}
The phase information of a nominal unlensed signal at the same band and signal-to-noise ratio $\rho_0$ is
\begin{equation}
I_0^{\rm ph}=\rho_0^2\frac{\int W(f)(f_0/f)^2\,\dd f}{\int W(f)\,\dd f},\qquad
\sigma_{\rm ideal}=\bigl(I_0^{\rm ph}\bigr)^{-1/2}.
\label{eq:idealinfo}
\end{equation}
Here $\sigma_{\rm ideal}$ refers to a single identifiable phenomenological coefficient of $f_0/f$, with the other parameters held fixed; it is not a coupling uncertainty. In a one-image model with two freely varying same-power coefficients $\gamma_{-1}$ and $\beta$, the Fisher matrix has rank one and the individual $\beta$ variance is undefined. With an independent Gaussian prior of width $\sigma_\gamma$ on the common source coefficient, the marginal single-image phase uncertainty is
\begin{equation}
\sigma_{\beta,{\rm single}}^2=\sigma_{\rm ideal}^2+\sigma_\gamma^2.
\label{eq:priorphase}
\end{equation}
For two point-lens images, $I_j=|\mu_j|I_0^{\rm ph}$, and the formal differential-phase error becomes
\begin{equation}
\frac{\sigma_{\Delta\beta,{\rm pair}}}{\sigma_{\rm ideal}}
=\sqrt{\frac1{|\mu_+|}+\frac1{|\mu_-|}}
=2.59002\quad(y=1).
\label{eq:penalty}
\end{equation}
Thus a pair does not automatically beat an ideally identifiable unlensed phase measurement in coefficient precision. Conversely, for an unrestricted common $1/f$ source contribution it provides a finite difference estimate where one image cannot. In the restricted two-coefficient comparison, the crossover occurs when $\sigma_\gamma/\sigma_{\rm ideal}=\sqrt{2.59002^2-1}=2.3892$. The same algebra applies to both detector bands because both images use the same normalized spectral shape. The two observables $\beta$ and $\Delta\beta$ are not the same model-dependent coupling: this crossover is a statistical information diagnostic only, not a claim of observational competitiveness for modified gravity.

\begin{table}[!t]
\centering\footnotesize
\setlength{\tabcolsep}{3.2pt}
\renewcommand{\arraystretch}{1.10}
\caption{Ideal phase errors and the source-prior crossover in the reduced model at $y=1$. Columns refer, respectively, to the single-image error, image-difference error, and crossover prior width; none constrains $Q$.}\label{tab:comparison}
\begin{tabular}{@{}lrrr@{}}\toprule
 & $\sigma_{\rm ideal}$ & $\sigma_{\Delta\beta}$ & $\sigma_{\gamma,\rm cross}$\\\midrule
ET   & $1.67\!\times\!10^{-3}$ & $4.33\!\times\!10^{-3}$ & $3.99\!\times\!10^{-3}$\\
LISA & $8.73\!\times\!10^{-7}$ & $2.26\!\times\!10^{-6}$ & $2.09\!\times\!10^{-6}$\\
\bottomrule\end{tabular}
\end{table}

The resulting comparison is given in Table~\ref{tab:comparison}.\par
\subsection{Reduced conditional waveform experiment}\label{sec:fisher5}
We use the analytical ET-D and LISA noise prescriptions and the standard point-lens magnification formula. The parameter errors are calculated from derivatives of the specified waveform and noise spectra. This is a reduced conditional calculation, not an event forecast or a fit to strain data.

For each resolved image $j=\pm$, after removing a known standard-GR lens transfer phase, we use
\begin{equation}
\begin{aligned}
\tilde h_j(f)={}&\mathcal A
\frac{\mathcal M_z^{5/6}f^{-7/6}}{D_{\rm lum}}
\sqrt{|\mu_j(y)|}\,e^{i\Psi_j(f)},\\
\Psi_j(f)={}&\Psi_N(f;\mathcal M_z)
+\left(c_{\rm com}+\frac{s_j\Delta\beta}{2}\right)
\frac{f_0}{f}.
\end{aligned}
\label{eq:fivewave}
\end{equation}
Here $s_+=1$, $s_-=-1$, and $f_0=10\,$Hz. The Newtonian phase is
\begin{equation*}
\Psi_N=\frac{3}{128}\left(\frac{\pi G_N\mathcal M_z f}{c^3}\right)^{-5/3}.
\end{equation*}
The nuisance parameter set is
\begin{equation}
\boldsymbol\theta=(\ln D_{\rm lum},\ln\mathcal M_z,c_{\rm com},\Delta\beta,y).
\end{equation}
The image magnifications are
\begin{equation*}
\mu_\pm=\frac12\pm\frac{y^2+2}{2y\sqrt{y^2+4}},
\end{equation*}
with their absolute values used for amplitudes. In this experiment $y$ affects the image amplitudes, but the time-delay and Morse-phase derivatives are excluded, as are sky position, inclination, mass ratio, spins, coalescence time, coalescence phase, the full PN/IMR structure, image-identification uncertainty, and detector evolution between arrival times. These omissions must be held fixed when interpreting the covariance.

Let $W(f)=f^{-7/3}/S_n(f)$ and $I_0=\int_{f_{\min}}^{f_{\max}}W(f)\,df$. Normalize the unlensed signal to the nominal high-SNR value $\rho_0$ and define $d_{aj}(f)=\partial_{\theta_a}\ln\tilde h_j$. Then the independently reconstructed Fisher elements are
\begin{equation}
\Gamma_{ab}=\sum_{j=\pm}\rho_0^2|\mu_j|\frac{1}{I_0}
\int_{f_{\min}}^{f_{\max}}W(f)\,
\mathrm{Re}[d_{aj}(f)d^*_{bj}(f)]\,df,
\label{eq:fivefisher}
\end{equation}
where the two image-noise realizations are taken as independent. The derivatives used in the code are
\begin{align}
&d_{\ln D_{\rm lum},j}=-1,\qquad
 d_{\ln\mathcal M_z,j}=\frac56-\frac{5i}{3}\Psi_N,\\
&d_{c_{\rm com},j}=i\frac{f_0}{f},\qquad
 d_{\Delta\beta,j}=\frac{i s_j}{2}\frac{f_0}{f},\\
&d_{y,j}=\frac12\partial_y\ln|\mu_j|.
\label{eq:derivatives}
\end{align}
The detector-frame inspiral cutoff follows the redshifted total binary mass, not the chirp mass. We use $f_{\min}=1\,$Hz and $f_{\max}=24.43\,$Hz for a $30+30M_\odot$ source-frame equal-mass binary at $z_s=2$ (so $\mathcal M_z=78.35M_\odot$), and $f_{\min}=10^{-4}\,$Hz, $f_{\max}=7.328\times10^{-4}\,$Hz for a $10^6+10^6M_\odot$ source-frame system at the same redshift ($\mathcal M_z=2.61165\times10^6 M_\odot$). The nominal unlensed SNRs $\rho_0=800$ (ET) and $50$ (LISA) are imposed diagnostic normalizations, not rates or guaranteed luminosity-distance predictions. The PSD prescriptions are the analytical fits cited in \citep{Hild2011,Robson2019} and reproduced in the code.

\begin{table*}[!t]
\centering\small
\setlength{\tabcolsep}{6pt}
\renewcommand{\arraystretch}{1.12}
\caption{Marginalized standard errors from the five-parameter Fisher matrix of Eq.~\eqref{eq:fivewave}, without a $y$ prior. Values refer to the reduced waveform experiment.}
\label{tab:fisher5}
\begin{tabular*}{0.91\textwidth}{@{\extracolsep{\fill}}lccccc@{}}\toprule
Setup & $\sigma_{\ln D_{\rm lum}}$ & $\sigma_{\ln\mathcal M_z}$ & $\sigma_{c_{\rm com}}$ & $\sigma_{\Delta\beta}$ & $\sigma_y$ \\\midrule
ET & $1.4481\times10^{-3}$ & $3.3535\times10^{-5}$ & $3.2610\times10^{-3}$ & $4.3263\times10^{-3}$ & $3.6197\times10^{-3}$ \\
LISA & $2.3191\times10^{-2}$ & $1.3003\times10^{-3}$ & $2.4735\times10^{-6}$ & $2.2610\times10^{-6}$ & $5.7915\times10^{-2}$ \\
\bottomrule\end{tabular*}
\end{table*}

The explicitly inverted five-parameter results are given in Table~\ref{tab:fisher5}. For LISA, an additional Gaussian $y$ prior of standard deviation $0.05$ or $0.02$ changes $(\sigma_{\ln D_{\rm lum}},\sigma_y)$ from $(0.02319,0.05791)$ to $(0.02003,0.03785)$ or $(0.01802,0.01890)$, respectively, in the same reduced likelihood. The differential-phase uncertainty remains $2.2610\times10^{-6}$ because in this factorized model the amplitude-only $y$ direction has no direct phase-gradient overlap. This numerical orthogonality is not generic when the lens time delay, finite-dimensional lens mass/profile, detector motion and source-model uncertainties are included.

The analytic two-image relation in equation~\eqref{eq:variance} reproduces the numerical $\sigma_{\Delta\beta}$ to the stated precision for this separable waveform. This provides a useful internal test of the calculation but does not elevate the conditional error to an astrophysical forecast. At the parameter level, a single event with freely varying common $1/f$ generation and propagation amplitudes remains rank deficient; observing a resolved pair identifies the differential coefficient, but a model-specific unlensed-versus-lensed coupling comparison additionally requires an independently derived binary-generation map, which is absent from the current action-level analysis.

\paragraph{Arrival-time and constant-phase nuisances.}\label{sec:timemorse}
The preceding five-parameter diagnostic assumes that the standard GR transfer phases, including the two geometric arrival times and image parity factors, can be removed without uncertainty. This is intentionally optimistic. Even when a lens is firmly identified, the measured signal contains the arrival-time phase $2\pi f t_j$ and a constant phase associated with coalescence and image Morse index. The Morse indices of standard geometric-optics images are discrete and may be determined by a trusted lens model; to quantify a deliberately conservative error budget, here we instead allow an independent continuous constant phase $\varphi_j$ and an independent arrival-time parameter $t_j$ for each image. This is a nuisance stress test, not the default physically optimal lens likelihood.

For this calculation extend the argument of the waveform in equation~\eqref{eq:fivewave} by
\begin{equation}
\Psi_j(f)\longmapsto\Psi_j(f)+\varphi_j+2\pi f t_j,\qquad j=\pm,
\label{eq:phasenui}
\end{equation}
and introduce the nine parameters
\begin{equation*}
\begin{aligned}
\boldsymbol\theta_9=(&\ln D_{\rm lum},\ln\mathcal M_z,
 c_{\rm com},\Delta\beta,y,\\
 &\varphi_+,\varphi_-,\tau_+,\tau_-),
\end{aligned}
\end{equation*}
where $\tau_j=2\pi f_{\rm piv}t_j$ and
$f_{\rm piv}=\sqrt{f_{\min}f_{\max}}$.
The new derivatives are
$\partial_{\varphi_j}\ln\tilde h_j=i$ and
$\partial_{\tau_j}\ln\tilde h_j=if/f_{\rm piv}$,
with zero contribution to the opposite image. Source and lens parameters not appearing in the reduced waveform remain fixed. The normalizations of the two images and their frequency intervals are unchanged, and this calculation does not impose a time-delay relation or a lens-geometry prior on $t_+-t_-$.

The degradation admits an independent moment-level derivation.
Let the normalized spectral weight be $p(f)\propto W(f)$,
with $\int p(f)\dd f=1$, and define
$\langle a,b\rangle=\int p(f)a(f)b(f)\dd f$.
Set $q(f)=f_0/f$, $u_0(f)=1$ and $u_1(f)=f/f_{\rm piv}$. Then
\begin{equation}
\begin{aligned}
B_{ab}&=\langle u_a,u_b\rangle,\\
v_a&=\langle q,u_a\rangle,\\
J_{\perp}&=\langle q,q\rangle
-v^\mathsf{T}B^{-1}v>0,
\quad a,b\in\{0,1\}.
\end{aligned}
\label{eq:projmoment}
\end{equation}
$J_{\perp}$ is the squared norm of the $f^{-1}$ phase direction after orthogonal projection away from the constant- and linear-frequency directions. For separable, identical normalized spectral shapes in the two images, their independent phase nuisances lead to
\begin{equation}
 \sigma_{\Delta\beta,{\rm free}\,\varphi,t}^{2}
 =\frac{1}{\rho_0^2 J_{\perp}}
 \left(\frac{1}{|\mu_+|}+\frac{1}{|\mu_-|}\right).
 \label{eq:projvar}
\end{equation}
Replacing the two-vector $(u_0,u_1)$ by $u_0$ alone gives the free-constant-phase result; omitting both nuisance directions returns equation~\eqref{eq:variance}. Equation~\eqref{eq:projvar} is valid for the defined conditional phase experiment, not a model-specific constraint on the covariant coupling $Q$. The additional common chirp-mass phase in the nine-parameter calculation is shared between both images; the numerical inversion below verifies that it does not change the differential-mode result under this particular waveform factorization.

\begin{table}[!t]
\centering\footnotesize
\setlength{\tabcolsep}{4.2pt}
\renewcommand{\arraystretch}{1.10}
\caption{Marginalized difference-phase errors with image-specific phase ($\varphi_\pm$) and arrival-time ($t_\pm$) nuisances. Ratio is relative to the five-parameter result.}\label{tab:phase_nuisance}
\begin{tabular}{@{}lrrrr@{}}\toprule
 & 5 par. & \shortstack{Phase\\free} & \shortstack{Phase/time\\free} & Ratio\\\midrule
ET & $4.33\cdot10^{-3}$ & $8.81\cdot10^{-3}$ & $1.48\cdot10^{-2}$ & 3.43\\
LISA & $2.26\cdot10^{-6}$ & $5.91\cdot10^{-6}$ & $1.35\cdot10^{-5}$ & 5.96\\
\bottomrule\end{tabular}
\end{table}

Table~\ref{tab:phase_nuisance} shows the two-detector nuisance comparison. For each detector, the $9\times9$ Fisher inverse and the independent Gram-matrix projection agree at numerical precision, and the frequency-integral results are stable under grid refinement. The two-sided nominal Gaussian $95\%$ phase thresholds accordingly change from $8.48\times10^{-3}$ to $2.90\times10^{-2}$ (ET) and from $4.43\times10^{-6}$ to $2.64\times10^{-5}$ (LISA). These thresholds concern the phenomenological differential phase only. In a physical analysis, the GR image delays, parity hypotheses, detector responses at separated epochs, intrinsic waveform parameters and the source/lens priors would have to enter a common likelihood. The calculation quantifies the sensitivity loss caused by freeing these nuisance directions within the stated Fisher model.

\begin{figure*}[!t]\centering
\includegraphics[width=.62\textwidth]{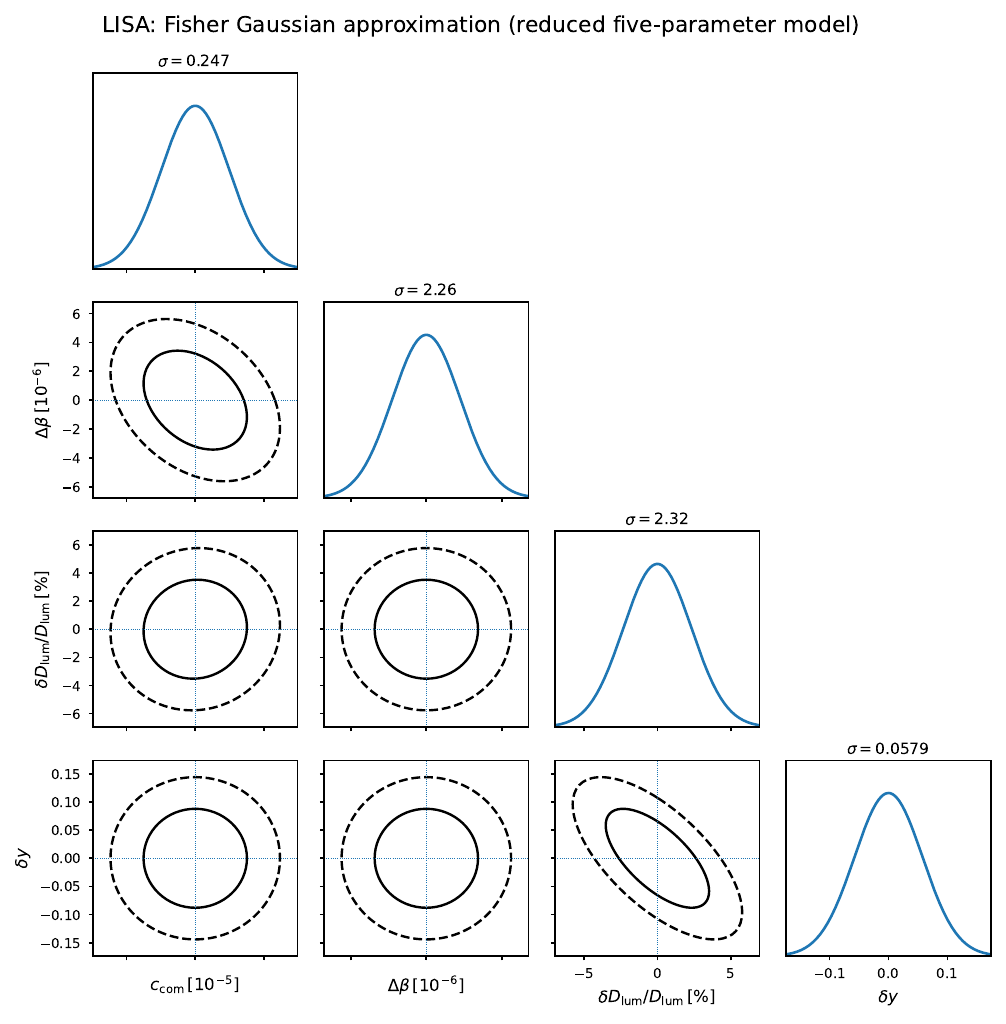}
\caption{Reduced five-parameter LISA Fisher covariance: marginalized Gaussian densities (diagonal) and 68\%/95\% contours (off-diagonal), with $\ln\mathcal M_z$ marginalized.}\label{fig:fishercorner}
\end{figure*}

Figure~\ref{fig:fishercorner} presents an analytic two-dimensional Gaussian summary of the recomputed Fisher covariance. Its elliptical contours express the reduced local likelihood approximation and are not numerical posterior samples.

These Fisher errors are conditioned on having both images available. The observing-window calculation in the following section tests that prerequisite separately from the formal phase precision.

\section{Time-delay selection and observing windows}\label{sec:timedelay}
For the standard point lens \citep{Takahashi2003,Bartelmann2010} and source position $y>0$,
\begin{equation}
x_\pm=\frac{y\pm\sqrt{y^2+4}}2,\qquad b_\pm=R_E|x_\pm|,
\end{equation}
The Einstein radius, in terms of angular-diameter distances, is
\begin{equation*}
R_E^2=\frac{4G_NM_LD_{\rm OL}D_{\rm LS}}{c^2D_{\rm OS}}.
\end{equation*}
The GR-reference observer-frame image time delay is
\begin{equation}
\begin{aligned}
\Delta t={}&\frac{4G_NM_L(1+z_L)}{c^3}\,\mathcal T(y),\\
\mathcal T(y)={}&\frac{y\sqrt{y^2+4}}2\\
&+\ln\!\left(\frac{\sqrt{y^2+4}+y}{\sqrt{y^2+4}-y}\right).
\end{aligned}
\label{eq:delay}
\end{equation}
At $z_L=0.5$, $z_s=2$, $y=1$, and $M_L=10^{14}\Msun$, the GR-reference delay is $194.84$ years (using $M_\odot=1.98847\times10^{30}$ kg and the standard physical constants). Both images would not be recorded in a four-year mission. A two-image Fisher diagnostic for this particular lens is therefore only a hypothetical both-images-available sensitivity, not a realistic event forecast.

\subsection{Finite observing-window envelope}\label{sec:window}
An explicit observing-window restriction can be imposed without an event-rate model. Write the positive point-lens delay function as
\begin{equation}
 {\cal T}(y)=\frac{y\sqrt{y^2+4}}2+\ln\left(\frac{\sqrt{y^2+4}+y}{\sqrt{y^2+4}-y}\right).
\end{equation}
Using the GR-reference geometry, requiring the two images to arrive within a window $T_{\rm win}$ yields the necessary point-lens condition
\begin{equation}
 M_L\le M_{\rm win}(y)=\frac{c^3T_{\rm win}}{4G_N(1+z_L){\cal T}(y)}.
 \label{eq:masswindow}
\end{equation}
Under the assumptions of equation~\eqref{eq:pndelay}, the PN-corrected mass-window bound is the GR bound multiplied by $2/(1+\gamma_{\rm PPN})$; for $|\gamma_{\rm PPN}-1|\le10^{-4}$ this differs by no more than approximately $5.01\times10^{-5}$ relative to the GR-reference mass window. For fixed lens and source redshifts and fixed $y$, the Einstein radius scales as $R_E\propto M_L^{1/2}$ and each surrogate differential phase has $|\Delta\beta_d|=C_d(M_L)\kappa_Q$ with $C_d\propto M_L^{1/2}$. Combining this GR-reference scaling with equation~\eqref{eq:masswindow} gives the surrogate-only, GR-reference window-conditioned inequality
\begin{equation}
\begin{aligned}
|\Delta\beta_d|&\le
C_d(M_{\rm win})\,\epsilon_{\rm PPN}^2/16,\\
&|1-\gamma_{\rm PPN}|\le\epsilon_{\rm PPN},
\quad d=1\mathrm D,3\mathrm D.
\end{aligned}
\label{eq:windowbound}
\end{equation}
At $z_L=0.5$, $z_s=2$, $y=1$ and $T_{\rm win}=4$ years, equation~\eqref{eq:masswindow} gives $M_{\rm win}=2.0530\times10^{12}\Msun$. For the illustrative $\epsilon_{\rm PPN}=10^{-4}$, the one-dimensional bound is $3.37\times10^{-34}$ and the distinct three-dimensional scalar-surrogate bound is $1.35\times10^{-33}$, approximately $5.95\times10^{-28}$ times the formal LISA one-sigma phase uncertainty. Using the PN time delay and PN impact parameter together would multiply the quoted GR-reference envelope by $[2/(1+\gamma_{\rm PPN})]^2$ at the corresponding limiting mass; the illustrative PPN criterion changes this number by at most about $1.01\times10^{-4}$. Neither this necessary time-window condition nor the fixed-SNR Fisher model predicts an astrophysical event rate or a complete tensor-mode phase. The bound illustrates the tradeoff imposed by observing the delayed image, and it would need to be recomputed using the full covariant propagation response before being used as a modified-gravity constraint.

\paragraph{Continuous-window image-pair selection.}\label{sec:windowselection}
The mass cap in equation~\eqref{eq:masswindow} is necessary, not sufficient, to observe two short-lived images. With the first image uniformly distributed inside a single continuous observing window $[0,T_{\rm win}]$, and conditional on both magnifications exceeding threshold, the second image is also within the window with probability
\begin{equation}
 P_{\rm pair}=(1-\Delta t/T_{\rm win})_+,\quad (x)_+\equiv\max(x,0).\label{eq:pairprob}
\end{equation}
For fixed $y$ and $z_L$, $\Delta t\propto M_L$. Setting $x=M_L/M_{\rm win}$ gives $P_{\rm pair}=(1-x)_+$. A scalar-surrogate information proxy, which assumes mass-independent conditional Fisher variance and $|\Delta\beta|\propto\sqrt{M_L}$, is
\begin{equation}
\begin{aligned}
{\cal W}(x)&=x(1-x),\quad 0\le x\le1,\\
{\cal W}_{\max}&=1/4\quad(x=1/2).
\end{aligned}\label{eq:selectionweight}
\end{equation}
For $T_{\rm win}=4$ yr, $y=1$ and $z_L=0.5$, $M_{\rm win}=2.0530\times10^{12}M_\odot$. A mass at the boundary has zero pair probability in this idealized short-signal model; half this mass gives a 2 yr delay and $P_{\rm pair}=1/2$. Equation~\eqref{eq:selectionweight} is neither an event rate nor an information forecast for the physical tensor coupling. It excludes a source population, detector duty cycle and long-duration waveforms \citep{LiMao2018}.

\section{Discussion and conclusions}\label{sec:discussion}
The static Einstein--scalar exterior has a genuine odd-parity tensor master equation, not a prescribed scalar flux surrogate \citep{Karlovini2002,ChowdhuryBanerjee2020}. Matching the exterior scalar charge to the conformal coupling and the local Newton constant yields Eq.~\eqref{eq:odd_delta}. The independently verified next-order result, Eq.~\eqref{eq:axialnextorder}, exposes a multipole-dependent coefficient and a sign change between the quadrupole and higher modes, providing an additional analytic check of the exact tensor potential. Equations~\eqref{eq:oddppnrelation} and \eqref{eq:oddGNmatched} make explicit how the radial order of the difference depends on whether the Schwarzschild baseline is matched at fixed Einstein-frame mass or fixed locally inferred matter mass. Its local $O(Q^2)$ term does not imply an observable $O(Q^2)$ inter-image phase. At fixed $\ell$, the Jacobian must accompany the radial WKB action; at fixed impact parameter the high-$\ell$ limit instead contains an order-$\omega$ geometric term and a distinct subleading annular correction. The FJNW-minus-Schwarzschild bending benchmark is achromatic. None of these finite-radius calculations includes the turning-point matching and wave-packet summation necessary for a lens transfer function.

The local scalar coupling also changes the weak-field lens map. Its point-mass achromatic observables constrain $M_L(1+\gamma_{\rm PPN})/2$, not $M_L$ and $Q$ separately; external mass information is needed. The constant-scalar vacuum black-hole source branch is GR-like, so the freely varying common source $1/f$ coefficient in the illustrative Fisher calculation is a stress-test parameter rather than a source prediction. Once independent image arrival times and constant phases are released, the conditional difference-phase uncertainty grows by factors $3.43$ (ET) and $5.96$ (LISA). With $y=1$ and $z_L=0.5$, a $10^{14}M_\odot$ point lens has a delay of approximately 195 yr, inconsistent with recording a resolved pair in a four-year window.

To construct a physical, GR-subtracted two-image dispersion signal one must solve the even-parity scalar--metric system, match a realistic matter-lens interior, implement uniform turning-point matching and multi-$\ell$ wave-packet propagation, and fit generation and propagation within the same action. This paper establishes narrower exterior and identifiability checks, not a completed observational bound on Horndeski couplings. The accompanying conditional $1/f$ illustration is therefore interpreted strictly as an information-geometry benchmark.

\section*{Acknowledgments}
This work was supported by the National Natural Science Foundation of China (Grant Nos. 12288102, 12573052 and 12475046) and by the Major Science and Technology Special Project of Xinjiang Uygur Autonomous Region (Grant No. 2022A03013-3).

\section*{Data and code availability}
The source and verification package accompanying this manuscript contains the scripts and numerical tables for the axial potential, radial and fixed-impact WKB comparisons, FJNW null-geodesic deflection, reduced Fisher matrices and finite-window selection. It also contains an independent symbolic check of the mass-normalization identities. The data are theoretical diagnostics rather than observed-event measurements. A public persistent repository link has not been assigned.
\appendix

\section{The local Newton constant in Brans--Dicke theory}\label{app:bd}
For completeness consider the Jordan-frame Brans--Dicke action \citep{BransDicke1961}
\begin{equation}
S=\frac{1}{16\pi G_*}\int\dd^4x\sqrt{-g}\left[\phi R-\frac{\omega_{\rm BD}}\phi(\nabla\phi)^2\right]+S_m.
\end{equation}
The field equations obtained by varying the Jordan-frame action are
\begin{align}
\phi G_{\mu\nu}={}&\frac{8\pi G_*}{c^4}T_{\mu\nu}
+\frac{\omega_{\rm BD}}{\phi}
\nabla_\mu\phi\nabla_\nu\phi\notag\\
&-\frac{\omega_{\rm BD}}{2\phi}
g_{\mu\nu}(\nabla\phi)^2\notag\\
&+\nabla_\mu\nabla_\nu\phi
-g_{\mu\nu}\Box\phi,\\
(2\omega_{\rm BD}+3)\Box\phi
={}&\frac{8\pi G_*}{c^4}T.
\label{eq:BDfield}
\end{align}
The scalar equation follows from the scalar variation combined with the trace of the metric equation for constant $\omega_{\rm BD}$. Both scalar and tensor perturbations contribute to the locally measured gravitational force, so $G_N$ cannot be identified solely with $G_*/\bar\phi$.

Writing $\phi=\bar\phi+\delta\phi$ and $g_{00}=-1+2U/c^2$ about a constant background, the static weak-field equations for a nonrelativistic source $\rho$ take the form
\begin{align}
\nabla^2\delta\phi&=-\frac{8\pi G_*}{(2\omega_{\rm BD}+3)c^2}\rho,\\
\nabla^2\left(U-\frac{c^2\delta\phi}{2\bar\phi}\right)&=-\frac{4\pi G_*}{\bar\phi}\rho.
\end{align}
Here the scalar equation and the appropriate linearized metric combination follow from the action, after suppressing time derivatives and higher weak-field terms. For a point mass, using $\nabla^2(1/r)=-4\pi\delta^{(3)}(\bm x)$, one obtains
\begin{equation}
\delta\phi=\frac{2G_*M}{(2\omega_{\rm BD}+3)c^2r},\quad
U=\frac{G_*M}{\bar\phi r}\left(1+\frac{1}{2\omega_{\rm BD}+3}\right).
\end{equation}
Comparing $U=G_NM/r$ gives
\begin{equation}
G_N=\frac{G_*}{\bar\phi}\frac{2\omega_{\rm BD}+4}{2\omega_{\rm BD}+3}.
\end{equation}
The local scalar charge is $\delta\phi/\bar\phi=q_L(G_NM/c^2r)$ with $q_L=(\omega_{\rm BD}+2)^{-1}$. This is a weak-field normalization identity, not a derivation of the lens-induced bGO kernel of Menadeo and Zumalac\'{a}rregui \citep{Menadeo2025}; the latter requires a matched coupled wave equation and the same GR-subtracted observable.

For an independent normalization check, define $\Phi=e^{2Q\phi/\mpl}$ in the conformal action~\eqref{eq:model}. Its canonical Jordan-frame kinetic density becomes
\begin{equation}
X=-\frac{\mpl^2}{8Q^2\Phi^2}(\nabla\Phi)^2,
\qquad \omega(\Phi)=\frac{1}{4Q^2\Phi}.
\end{equation}
Thus this model is a variable-$\omega$ scalar--tensor theory, not globally the constant-$\omega_{\rm BD}$ model. At the local asymptotic background $\Phi_0=1$ the effective weak-field parameter is $\omega_0=1/(4Q^2)$, giving
\begin{equation}
\gamma_{\rm PPN}=\frac{\omega_0+1}{\omega_0+2}
=\frac{1+4Q^2}{1+8Q^2},
\end{equation}
in agreement with equation~\eqref{eq:gammaPPN}. These identities require a massless unscreened scalar and weakly self-gravitating sources; they are not automatically valid for screened or strongly self-gravitating objects.

\section{Horndeski notation and the shortwave hierarchy}\label{app:horndeski}
For reference, the four familiar covariant Horndeski densities \citep{Kobayashi2011,Bettoni2013} in our conventions are
\begin{align}
\mathcal L_2&=G_2(\phi,X),\\
\mathcal L_3&=-G_3(\phi,X)\Box\phi,\\
\mathcal L_4&=G_4R+G_{4X}\left[(\Box\phi)^2-\phi_{;\mu\nu}\phi^{;\mu\nu}\right],\\
\mathcal L_5={}&G_5G_{\mu\nu}\phi^{;\mu\nu}
-\frac{G_{5X}}{6}(\Box\phi)^3\notag\\
&+\frac{G_{5X}}{2}\Box\phi\,
\phi_{;\mu\nu}\phi^{;\mu\nu}\notag\\
&-\frac{G_{5X}}{3}\,
\phi_{;\mu}{}^\nu\phi_{;\nu}{}^\rho\phi_{;\rho}{}^\mu.
\end{align}
The background quantities $\alpha_K$ and $\alpha_B$ are standard FLRW projections and, in general, are insufficient to determine the local principal symbol \eqref{eq:fullop}. When the coefficients change on a scale much longer than the wavelength, write $\psi^I=e^{i\theta/\epsilon}\sum_{n\ge0}\epsilon^n a_n^I$. The leading equation is $K_{IJ}^{\mu\nu}k_\mu k_\nu a_0^J=0$; the next orders involve derivatives of the same background operator and of the polarization eigenvectors. For a nonluminal branch, the correct characteristic need not coincide with the metric light cone. The diagonal, luminal, one-dimensional reduction \eqref{eq:oneD} is an additional approximation, not a consequence of replacing the full Horndeski principal symbol with the metric null symbol. A complete calculation must recover \eqref{eq:oneD} from the action or supplement it with the omitted same-order terms.

{\footnotesize\setlength{\bibsep}{2pt}

}


\begin{thebibliography}{47}
\providecommand{\natexlab}[1]{#1}
\providecommand{\url}[1]{\texttt{#1}}
\expandafter\ifx\csname urlstyle\endcsname\relax
  \providecommand{\doi}[1]{doi: #1}\else
  \providecommand{\doi}{doi: \begingroup \urlstyle{rm}\Url}\fi

\bibitem[{Abbott} et~al.(2016){Abbott}, {Abbott}, {Abbott}, {Abernathy},
  {Acernese}, et~al.]{Abbott2016}
B.~P. {Abbott}, R.~{Abbott}, T.~D. {Abbott}, M.~R. {Abernathy}, F.~{Acernese},
  et~al.
\newblock {Tests of General Relativity with GW150914}.
\newblock \emph{Physical Review Letters}, 116\penalty0 (22):\penalty0 221101,
  July 2016.
\newblock \doi{10.1103/PhysRevLett.116.221101}.

\bibitem[{Abbott} et~al.(2019){Abbott}, {Abbott}, {Abbott},
  et~al.]{Abbott2019PRL}
B.~P. {Abbott}, R.~{Abbott}, T.~D. {Abbott}, et~al.
\newblock {Tests of General Relativity with GW170817}.
\newblock \emph{Physical Review Letters}, 123\penalty0 (1):\penalty0 011102,
  2019.
\newblock \doi{10.1103/PhysRevLett.123.011102}.

\bibitem[{Abbott} et~al.(2021){Abbott}, {Abbott}, {Abraham},
  et~al.]{Abbott2021}
R.~{Abbott}, T.~D. {Abbott}, S.~{Abraham}, et~al.
\newblock {Tests of General Relativity with Binary Black Holes from the second
  LIGO-Virgo Gravitational-Wave Transient Catalog}.
\newblock \emph{Physical Review D}, 103\penalty0 (12):\penalty0 122002, 2021.
\newblock \doi{10.1103/PhysRevD.103.122002}.

\bibitem[{Abbott} et~al.(2017){Abbott}, {Abbott}, {Abbott}, et~al.]{Abbott2017}
B.~P. {Abbott}, R.~{Abbott}, T.~D. {Abbott}, et~al.
\newblock {GW170817: Observation of Gravitational Waves from a Binary Neutron
  Star Inspiral}.
\newblock \emph{Physical Review Letters}, 119\penalty0 (16):\penalty0 161101,
  2017.
\newblock \doi{10.1103/PhysRevLett.119.161101}.

\bibitem[{Ezquiaga} and {Zumalac{\'a}rregui}(2017)]{Ezquiaga2017}
J.~M. {Ezquiaga} and M.~{Zumalac{\'a}rregui}.
\newblock {Dark Energy After GW170817: Dead Ends and the Road Ahead}.
\newblock \emph{Physical Review Letters}, 119\penalty0 (25):\penalty0 251304,
  2017.
\newblock \doi{10.1103/PhysRevLett.119.251304}.

\bibitem[Brans and Dicke(1961)]{BransDicke1961}
C.~Brans and R.~H. Dicke.
\newblock Mach's principle and a relativistic theory of gravitation.
\newblock \emph{Physical Review}, 124:\penalty0 925--935, 1961.
\newblock \doi{10.1103/PhysRev.124.925}.

\bibitem[{Horndeski}(1974)]{Horndeski1974}
G.~W. {Horndeski}.
\newblock {Second-Order Scalar-Tensor Field Equations in a Four-Dimensional
  Space}.
\newblock \emph{International Journal of Theoretical Physics}, 10\penalty0
  (6):\penalty0 363--384, 1974.
\newblock \doi{10.1007/BF01807638}.

\bibitem[{Clifton} et~al.(2012){Clifton}, {Ferreira}, {Padilla}, and
  {Skordis}]{Clifton2012}
T.~{Clifton}, P.~G. {Ferreira}, A.~{Padilla}, and C.~{Skordis}.
\newblock {Modified Gravity and Cosmology}.
\newblock \emph{Physics Reports}, 513\penalty0 (1-3):\penalty0 1--189, 2012.
\newblock \doi{10.1016/j.physrep.2012.01.001}.

\bibitem[{Kobayashi} et~al.(2011){Kobayashi}, {Yamaguchi}, and
  {Yokoyama}]{Kobayashi2011}
T.~{Kobayashi}, M.~{Yamaguchi}, and J.~{Yokoyama}.
\newblock {Generalized G-Inflation: Inflation with the Most General
  Second-Order Field Equations}.
\newblock \emph{Progress of Theoretical Physics}, 126\penalty0 (3):\penalty0
  511--529, 2011.
\newblock \doi{10.1143/PTP.126.511}.

\bibitem[{Bettoni} et~al.(2013){Bettoni}, {Liberati}, and
  {Sindoni}]{Bettoni2013}
D.~{Bettoni}, S.~{Liberati}, and L.~{Sindoni}.
\newblock {Extended Library of Horndeski Theories}.
\newblock \emph{Journal of Cosmology and Astroparticle Physics}, 2013\penalty0
  (11):\penalty0 007, 2013.
\newblock \doi{10.1088/1475-7516/2013/11/007}.

\bibitem[Bellini and Sawicki(2014)]{Bellini2014}
Emilio Bellini and Ignacy Sawicki.
\newblock Maximal freedom at minimum cost: Linear large-scale structure in
  general modifications of gravity.
\newblock \emph{Journal of Cosmology and Astroparticle Physics}, 2014\penalty0
  (07):\penalty0 050, 2014.
\newblock \doi{10.1088/1475-7516/2014/07/050}.

\bibitem[Saltas et~al.(2014)Saltas, Sawicki, Amendola, and Kunz]{Saltas2014}
Ippocratis~D. Saltas, Ignacy Sawicki, Luca Amendola, and Martin Kunz.
\newblock Anisotropic stress as a signature of nonstandard propagation of
  gravitational waves.
\newblock \emph{Physical Review Letters}, 113:\penalty0 191101, 2014.
\newblock \doi{10.1103/PhysRevLett.113.191101}.

\bibitem[Lombriser and Taylor(2016)]{Lombriser2016}
Lucas Lombriser and Andy Taylor.
\newblock Breaking a dark degeneracy with gravitational waves.
\newblock \emph{Journal of Cosmology and Astroparticle Physics}, 2016\penalty0
  (03):\penalty0 031, 2016.
\newblock \doi{10.1088/1475-7516/2016/03/031}.

\bibitem[Belgacem et~al.(2018)Belgacem, Dirian, Foffa, and
  Maggiore]{Belgacem2018}
Enis Belgacem, Yves Dirian, Stefano Foffa, and Michele Maggiore.
\newblock Modified gravitational-wave propagation and standard sirens.
\newblock \emph{Physical Review D}, 98:\penalty0 023510, 2018.
\newblock \doi{10.1103/PhysRevD.98.023510}.

\bibitem[Nishizawa(2018)]{Nishizawa2018}
Atsushi Nishizawa.
\newblock Generalized framework for testing gravity with gravitational-wave
  propagation. {I}. formulation.
\newblock \emph{Physical Review D}, 97:\penalty0 104037, 2018.
\newblock \doi{10.1103/PhysRevD.97.104037}.

\bibitem[Ezquiaga and Zumalac{\'a}rregui(2018)]{Ezquiaga2018Frontiers}
Jose~Maria Ezquiaga and Miguel Zumalac{\'a}rregui.
\newblock Dark energy in light of multi-messenger gravitational-wave astronomy.
\newblock \emph{Frontiers in Astronomy and Space Sciences}, 5:\penalty0 44,
  2018.
\newblock \doi{10.3389/fspas.2018.00044}.

\bibitem[{Takahashi} and {Nakamura}(2003)]{Takahashi2003}
R.~{Takahashi} and T.~{Nakamura}.
\newblock {Wave Effects in the Gravitational Lensing of Gravitational Waves
  from Chirping Binaries}.
\newblock \emph{The Astrophysical Journal}, 595\penalty0 (2):\penalty0
  1039--1051, 2003.
\newblock \doi{10.1086/377430}.

\bibitem[{Dai} et~al.(2017){Dai}, {Venumadhav}, and {Sigurdson}]{Dai2017}
L.~{Dai}, T.~{Venumadhav}, and K.~{Sigurdson}.
\newblock {Effect of Lens Mass Distribution on Gravitational Wave Signals}.
\newblock \emph{Physical Review D}, 95\penalty0 (4):\penalty0 044011, 2017.
\newblock \doi{10.1103/PhysRevD.95.044011}.

\bibitem[{Bartelmann}(2010)]{Bartelmann2010}
M.~{Bartelmann}.
\newblock {Gravitational Lensing}.
\newblock \emph{Classical and Quantum Gravity}, 27\penalty0 (23):\penalty0
  233001, 2010.
\newblock \doi{10.1088/0264-9381/27/23/233001}.

\bibitem[Ezquiaga and Zumalac{\'a}rregui(2020)]{Ezquiaga2020PRD}
Jose~Maria Ezquiaga and Miguel Zumalac{\'a}rregui.
\newblock Gravitational wave lensing beyond general relativity: Birefringence,
  echoes, and shadows.
\newblock \emph{Physical Review D}, 102\penalty0 (12):\penalty0 124048, 2020.
\newblock \doi{10.1103/PhysRevD.102.124048}.

\bibitem[Ezquiaga et~al.(2021)Ezquiaga, Holz, Hu, Lagos, and
  Wald]{Ezquiaga2021PRD}
Jose~Maria Ezquiaga, Daniel~E. Holz, Wayne Hu, Macarena Lagos, and Robert~M.
  Wald.
\newblock Phase effects from strong gravitational lensing of gravitational
  waves.
\newblock \emph{Physical Review D}, 103\penalty0 (6):\penalty0 064047, 2021.
\newblock \doi{10.1103/PhysRevD.103.064047}.

\bibitem[Goyal et~al.(2023)Goyal, Vijaykumar, Ezquiaga, and
  Zumalac{\'a}rregui]{Goyal2023PRD}
Srashti Goyal, Aditya Vijaykumar, Jose~Maria Ezquiaga, and Miguel
  Zumalac{\'a}rregui.
\newblock Probing lens-induced gravitational-wave birefringence as a test of
  general relativity.
\newblock \emph{Physical Review D}, 108\penalty0 (2):\penalty0 024052, 2023.
\newblock \doi{10.1103/PhysRevD.108.024052}.

\bibitem[{Savastano} et~al.(2019){Savastano}, {Hannuksela}, {Barausse}, and
  {Collett}]{Savastano2019PRD}
S.~{Savastano}, O.~A. {Hannuksela}, E.~{Barausse}, and T.~{Collett}.
\newblock {Gravitational-Wave Lensing of Binary Systems}.
\newblock \emph{Physical Review D}, 100\penalty0 (10):\penalty0 104035, 2019.
\newblock \doi{10.1103/PhysRevD.100.104035}.

\bibitem[{Hild} et~al.(2011){Hild}, {Abernathy}, {Acernese}, et~al.]{Hild2011}
S.~{Hild}, M.~{Abernathy}, F.~{Acernese}, et~al.
\newblock {Sensitivity Studies for Third-Generation Gravitational Wave
  Observatories}.
\newblock \emph{Classical and Quantum Gravity}, 28\penalty0 (9):\penalty0
  094013, 2011.
\newblock \doi{10.1088/0264-9381/28/9/094013}.

\bibitem[{Amaro-Seoane} et~al.(2017){Amaro-Seoane}, {Audley}, {Babak},
  et~al.]{AmaroSeoane2017}
P.~{Amaro-Seoane}, H.~{Audley}, S.~{Babak}, et~al.
\newblock {Laser Interferometer Space Antenna}.
\newblock \emph{arXiv e-prints}, art. arXiv:1702.00786, February 2017.

\bibitem[Robson et~al.(2019)Robson, Cornish, and Liu]{Robson2019}
Travis Robson, Neil~J. Cornish, and Chang Liu.
\newblock The construction and use of lisa sensitivity curves.
\newblock \emph{Class. Quant. Grav.}, 36\penalty0 (10):\penalty0 105011, 2019.

\newblock \doi{10.1088/1361-6382/ab1101}.

\bibitem[{Auclair} et~al.(2023){Auclair}, {Baker}, {Barausse},
  et~al.]{Auclair2023}
O.~{Auclair}, J.~{Baker}, E.~{Barausse}, et~al.
\newblock {Cosmology with the Laser Interferometer Space Antenna}.
\newblock \emph{Living Reviews in Relativity}, 26\penalty0 (1):\penalty0 5,
  2023.
\newblock \doi{10.1007/s41114-023-00045-2}.

\bibitem[Grespan and Biesiada(2023)]{Grespan2023}
Margherita Grespan and Marek Biesiada.
\newblock Strong gravitational lensing of gravitational waves: A review.
\newblock \emph{Universe}, 9\penalty0 (5):\penalty0 200, 2023.
\newblock \doi{10.3390/universe9050200}.

\bibitem[{Finke} et~al.(2021){Finke}, {Foffa}, {Iacovelli}, {Maggiore}, and
  {Mancarella}]{Finke2021}
Andreas {Finke}, Stefano {Foffa}, Francesco {Iacovelli}, Michele {Maggiore},
  and Michele {Mancarella}.
\newblock {Probing modified gravitational wave propagation with strongly lensed
  coalescing binaries}.
\newblock \emph{\prd}, 104\penalty0 (8):\penalty0 084057, October 2021.
\newblock \doi{10.1103/PhysRevD.104.084057}.

\bibitem[Li et~al.(2018)Li, Mao, Zhao, and Lu]{LiMao2018}
Shun-Sheng Li, Shude Mao, Yuetong Zhao, and Youjun Lu.
\newblock Gravitational lensing of gravitational waves: a statistical
  perspective.
\newblock \emph{Monthly Notices of the Royal Astronomical Society},
  476:\penalty0 2220--2229, 2018.
\newblock \doi{10.1093/mnras/sty411}.

\bibitem[{Isaacson}(1968)]{Isaacson1968}
R.~A. {Isaacson}.
\newblock {Gravitational Radiation in the Limit of High Frequency. I. The
  Linear Approximation and Geometrical Optics}.
\newblock \emph{Physical Review}, 166\penalty0 (5):\penalty0 1263--1271, 1968.
\newblock \doi{10.1103/PhysRev.166.1263}.

\bibitem[Dalang et~al.(2021)Dalang, Fleury, and Lombriser]{Dalang2021}
Charles Dalang, Pierre Fleury, and Lucas Lombriser.
\newblock Scalar and tensor gravitational waves.
\newblock \emph{Physical Review D}, 103:\penalty0 064075, 2021.
\newblock \doi{10.1103/PhysRevD.103.064075}.

\bibitem[Garoffolo et~al.(2020)Garoffolo, Tasinato, Carbone, Bertacca, and
  Matarrese]{Garoffolo2020}
Alice Garoffolo, Gianmassimo Tasinato, Carmelita Carbone, Daniele Bertacca, and
  Sabino Matarrese.
\newblock Gravitational waves and geometrical optics in scalar-tensor theories.
\newblock \emph{Journal of Cosmology and Astroparticle Physics}, 2020\penalty0
  (11):\penalty0 040, 2020.
\newblock \doi{10.1088/1475-7516/2020/11/040}.

\bibitem[{Menadeo} and {Zumalac{\'a}rregui}(2025)]{Menadeo2025}
N.~{Menadeo} and M.~{Zumalac{\'a}rregui}.
\newblock {Gravitational wave propagation beyond general relativity: Geometric
  optic expansion and lens-induced dispersion}.
\newblock \emph{Physical Review D}, 111\penalty0 (10):\penalty0 104022, 2025.
\newblock \doi{10.1103/PhysRevD.111.104022}.

\bibitem[Menadeo et~al.(2026)Menadeo, Giardino, and
  Zumalac{\\'a}rregui]{MenadeoGiardino2026}
Nicola Menadeo, Serena Giardino, and Miguel Zumalac{\\'a}rregui.
\newblock Gravitational-wave dispersion over inhomogeneous space-times: General
  relativity, screened theories of gravity and non-minimal dark energy.
\newblock \emph{Journal of Cosmology and Astroparticle Physics}, 2026\penalty0
  (09):\penalty0 092, 2026.
\newblock \doi{10.1088/1475-7516/2026/09/092}.

\bibitem[Garoffolo and Tasinato(2026)]{GaroffoloTasinato2026}
Alice Garoffolo and Gianmassimo Tasinato.
\newblock Wave-optics gravitational wave lensing in modified gravity.
\newblock \emph{Physical Review D}, 114:\penalty0 064059, 2026.
\newblock \doi{10.1103/smzt-5v1s}.

\bibitem[Nashed(2026)]{Nashed2026Odd}
G.~G.~L. Nashed.
\newblock Odd-parity perturbations of geometrically regular black holes with
  hedgehog scalar hair.
\newblock \emph{The European Physical Journal C}, 86:\penalty0 1289, 2026.
\newblock \doi{10.1140/epjc/s10052-026-16584-1}.

\bibitem[Lagos et~al.(2018)Lagos, Bellini, Noller, Ferreira, and
  Baker]{Lagos2018}
Macarena Lagos, Emilio Bellini, Johannes Noller, Pedro~G. Ferreira, and Tessa
  Baker.
\newblock A general theory of linear cosmological perturbations: stability
  conditions, the quasistatic limit and dynamics.
\newblock \emph{Journal of Cosmology and Astroparticle Physics}, 2018\penalty0
  (03):\penalty0 021, 2018.
\newblock \doi{10.1088/1475-7516/2018/03/021}.

\bibitem[Karlovini(2002)]{Karlovini2002}
Max Karlovini.
\newblock Axial perturbations of general spherically symmetric space-times.
\newblock \emph{Classical and Quantum Gravity}, 19:\penalty0 2125--2140, 2002.
\newblock \doi{10.1088/0264-9381/19/8/305}.

\bibitem[Chowdhury and Banerjee(2020)]{ChowdhuryBanerjee2020}
Avijit Chowdhury and Narayan Banerjee.
\newblock Echoes from a singularity.
\newblock \emph{Physical Review D}, 102:\penalty0 124051, 2020.
\newblock \doi{10.1103/PhysRevD.102.124051}.

\bibitem[Will(2014)]{Will2014}
Clifford~M. Will.
\newblock The confrontation between general relativity and experiment.
\newblock \emph{Living Reviews in Relativity}, 17:\penalty0 4, 2014.
\newblock \doi{10.12942/lrr-2014-4}.

\bibitem[{Cutler} and {Flanagan}(1994)]{Cutler1994}
C.~{Cutler} and E.~E. {Flanagan}.
\newblock {Gravitational Waves from Merging Compact Binaries: How Accurately
  Can One Extract the Binary's Parameters from the Inspiral Waveform?}
\newblock \emph{Physical Review D}, 49\penalty0 (6):\penalty0 2658--2697, 1994.
\newblock \doi{10.1103/PhysRevD.49.2658}.

\bibitem[Vallisneri(2008)]{Vallisneri2008}
Michele Vallisneri.
\newblock Use and abuse of the fisher information matrix in the assessment of
  gravitational-wave parameter-estimation prospects.
\newblock \emph{Physical Review D}, 77\penalty0 (4):\penalty0 042001, 2008.
\newblock \doi{10.1103/PhysRevD.77.042001}.

\bibitem[{Carson} et~al.(2020){Carson}, {Haster}, {Chatziioannou},
  et~al.]{Carson2020}
Z.~{Carson}, C.-J. {Haster}, K.~{Chatziioannou}, et~al.
\newblock {The Impact of Non-Tensorial Polarizations on the Detection and
  Characterization of Gravitational-Wave Signals}.
\newblock \emph{Classical and Quantum Gravity}, 37\penalty0 (15):\penalty0
  155005, 2020.
\newblock \doi{10.1088/1361-6382/ab9a1e}.

\bibitem[Perlick(2004)]{Perlick2004}
Volker Perlick.
\newblock Gravitational lensing from a spacetime perspective.
\newblock \emph{Living Reviews in Relativity}, 7:\penalty0 9, 2004.
\newblock \doi{10.12942/lrr-2004-9}.

\bibitem[Bertotti et~al.(2003)Bertotti, Iess, and Tortora]{Bertotti2003}
Bruno Bertotti, Luciano Iess, and Paolo Tortora.
\newblock A test of general relativity using radio links with the cassini
  spacecraft.
\newblock \emph{Nature}, 425:\penalty0 374--376, 2003.
\newblock \doi{10.1038/nature01997}.

\bibitem[Guo et~al.(2026)Guo, Cao, Wang, Chen, Biesiada, Liu, and
  Lian]{Guo2026LensingPPN}
Wuzheng Guo, Shuo Cao, Qiumin Wang, Yun Chen, Marek Biesiada, Tonghua Liu, and
  Yujie Lian.
\newblock Extragalactic test of general relativity with time-delay
  gravitational lenses.
\newblock \emph{European Physical Journal C}, 86:\penalty0 650, 2026.
\newblock \doi{10.1140/epjc/s10052-026-15856-0}.

\end{thebibliography}
\end{document}